\documentclass[journal]{IEEEtran}

\usepackage{amsmath,amssymb}
\usepackage{etoolbox}
\makeatletter
\def\@IEEEBIOskipN{4pt}
\makeatother
\BeforeBeginEnvironment{IEEEbiography}{\vskip 0pt plus -1fil\relax}
\usepackage{booktabs}
\usepackage{tabularx}

\usepackage{graphicx}
\usepackage{tikz}
\usetikzlibrary{arrows.meta,positioning,fit,backgrounds,decorations.pathreplacing}
\usepackage[edges]{forest}
\renewcommand{\arraystretch}{0.91}
\usepackage[numbers,sort&compress]{natbib}
\usepackage[colorlinks=true,linkcolor=blue,citecolor=blue,urlcolor=blue]{hyperref}

\begin{document}

\title{Dependency, Compression, and Synergy:\\
A Unified Information-Theoretic View of Multimodal Learning}

\author{Liangjian Wen, Linjie Li, Jiang Duan, Yong Dai, Jianzhuang Liu, Zhao Kang
\thanks{
L. Wen, L. Li, and J. Duan are with the School of Computing and Artificial Intelligence,
Southwestern University of Finance and Economics, Chengdu, China. E-mail: wlj6816@gmail.com
(Corresponding author: Liangjian Wen.)
}
\thanks{
Y. Dai is with the Beijing Humanoid Robot Innovation Center (X-Humanoid), Beijing, China.
}
\thanks{
J. Liu is with the Shenzhen Institutes of Advanced Technology,
Chinese Academy of Sciences, Shenzhen, China.
}
\thanks{
Z. Kang is with the University of Electronic Science and Technology of China,
Chengdu, China.
}
}

\markboth{IEEE Transactions on Pattern Analysis and Machine Intelligence}%
{Wen \MakeLowercase{\textit{et al.}}: A Unified Information-Theoretic View of Multimodal Learning}

\maketitle

\begin{abstract}
Recent advances in multimodal foundation models have intensified the need to understand how different modalities share, preserve, and complement information. Mutual Information (MI), the Information Bottleneck (IB), and Partial Information Decomposition (PID) provide complementary perspectives on this question, yet existing studies often treat them as isolated theoretical tools. This survey presents an information-theoretic perspective that connects these principles as progressively refined views of multimodal information processing: MI characterizes inter-modal dependency, IB explains task-oriented information preservation under compression, and PID further decomposes preserved information into redundancy, uniqueness, and synergy. We review 170 recent studies (2018--2026) together with 12 foundational works, organizing multimodal learning around four fundamental challenges: cross-modal alignment, information-efficient fusion, interaction-type characterization, and scaling to multimodal foundation models. Rather than using application domains as primary taxonomy axes, we interpret healthcare, robotics, recommendation systems, affective computing, and wireless communications as empirical validations of these information principles. Beyond taxonomy, we organize existing multimodal paradigms within a single information-theoretic coordinate system -- the Generalized Multimodal Information Lagrangian -- in which they occupy exact or approximate parameter corners, and whose unoccupied regions name candidate method families the literature has not yet built. We further discuss how emerging multimodal foundation models instantiate these principles at scale and identify open challenges, including scalable information estimation in high-dimensional settings, standardized evaluation across information-theoretic methods, combinatorial complexity of multimodal PID, and the transition from post-hoc information analysis toward information-aware multimodal learning.
\end{abstract}

\begin{IEEEkeywords}
Multimodal learning, mutual information, information bottleneck, partial information decomposition, representation learning, information theory, survey.
\end{IEEEkeywords}

\IEEEpeerreviewmaketitle
\bstctlcite{IEEEexample:BSTcontrol}

\section{Introduction}
\label{sec:intro}

\IEEEPARstart{M}{ultimodal} learning asks a model to make sense of the world from more than one channel of evidence at once -- pixels and words, audio and video, clinical images and health records, sensor streams and text. Modalities are complementary: what is ambiguous in one channel is often resolved by another. Realizing that promise is far harder than concatenating features or stacking cross-attention layers, and four challenges recur across nearly every surveyed application. First, \emph{cross-modal alignment}: how should encoders be trained so that semantically corresponding content -- an image and its caption, a robot's camera frame and a spoken instruction -- lands in a shared representational space, as popularized at scale by CLIP-style contrastive pretraining \citep{Radford2021Learning}? Second, \emph{robust, information-efficient fusion}: how should heterogeneous signals be combined so the result reflects genuinely joint evidence, survives a missing sensor or corrupted channel, and resists letting the fastest-converging modality dominate? Third, \emph{interaction-type characterization}: even fusion robust to noise and missingness cannot say what kind of cross-modal relationship does the predictive work -- whether two modalities merely repeat one another, one carries something the other lacks, or the prediction is possible only from their combination. Fourth, \emph{scaling to foundation models}: interpretability, efficiency, and diagnosis at scale. As multimodal systems are increasingly built by adapting a pretrained foundation model rather than training encoders from scratch, practitioners need tools to explain, prune, and diagnose a model whose representations were never designed with the first three challenges in mind.

\begin{figure*}[t]
\centering
\resizebox{\textwidth}{!}{%
\begin{forest}
  forked edges,
  for tree={
    grow=east, reversed=true,
    anchor=base west, parent anchor=east, child anchor=west, base=left,
    font=\footnotesize,
    rectangle, rounded corners=2pt, draw=black!55, thick,
    align=left, inner xsep=4pt, inner ysep=2pt,
    edge={draw=black!45, thick}, l sep=10pt, s sep=1.5pt,
  },
  where level=0{fill=black!8, align=center, font=\footnotesize\bfseries}{},
  where level=1{align=center, text width=2.75cm, font=\footnotesize\bfseries}{},
  where level=2{text width=10.6cm, font=\scriptsize}{},
  [{Information-Theoretic\\Multimodal Learning\\\mdseries\scriptsize three principles\\\mdseries\scriptsize about dependency}
    [{Aligning\\Representations\\(\S\ref{sec:align})\\\mdseries\scriptsize P1: dependency\\\mdseries\scriptsize is measurable}, fill=blue!16, draw=blue!55
      [{\textbf{Estimation toolbox} (\S\ref{sec:align-toolbox})\textbf{:} MINE, InfoNCE, CLUB, RINCE}, fill=blue!7, draw=blue!35]
      [{\textbf{CLIP-style MI maximization} (\S\ref{sec:align-clip})\textbf{:} CLIP, CLOOB, TupleInfoNCE, MGCA, LIV}, fill=blue!7, draw=blue!35]
      [{\textbf{IB-regularized alignment} (\S\ref{sec:align-ib})\textbf{:} CIBR, OVA-IB, GARE, DVSIB}, fill=blue!7, draw=blue!35]
      [{\textbf{PID-synergy contrastive} (\S\ref{sec:align-pid})\textbf{:} FactorCL, CoMM, InfMasking, DeCUR}, fill=blue!7, draw=blue!35]
    ]
    [{Compressing for\\Robust Fusion\\(\S\ref{sec:robust})\\\mdseries\scriptsize P2: it must\\\mdseries\scriptsize be allocated}, fill=orange!18, draw=orange!60
      [{\textbf{MI--IB hybrid fusion} (\S\ref{sec:robust-hybrid})\textbf{:} MVIB, MIB, MMIM, MuMMI}, fill=orange!8, draw=orange!40]
      [{\textbf{Imbalance \& shift} (\S\ref{sec:robust-imbalance})\textbf{:} OMIB, Elastic IB}, fill=orange!8, draw=orange!40]
      [{\textbf{Missing modalities} (\S\ref{sec:robust-missing})\textbf{:} CyIN, RedCore, I\textsuperscript{3}-MRec}, fill=orange!8, draw=orange!40]
      [{\textbf{Noise \& shortcuts} (\S\ref{sec:robust-noise})\textbf{:} MCIB, ITHP, CaMIB}, fill=orange!8, draw=orange!40]
      [{\textbf{Cross-modal distillation} (\S\ref{sec:robust-distillation})\textbf{:} AMID, complementarity criterion}, fill=orange!8, draw=orange!40]
      [{\textbf{Transmission \& extraction} (\S\ref{sec:robust-transmission})\textbf{:} Distributed IB, IBMEA, InfoMeter}, fill=orange!8, draw=orange!40]
    ]
    [{Decomposing \&\\Typing Interaction\\(\S\ref{sec:interaction})\\\mdseries\scriptsize P3: it is not\\\mdseries\scriptsize monolithic}, fill=green!18, draw=green!55
      [{\textbf{Pre-PID clustering} (\S\ref{sec:interaction-preformal})\textbf{:} SIB-MSC, DVIB, MRDD}, fill=green!7, draw=green!40]
      [{\textbf{PID theory \& estimators} (\S\ref{sec:interaction-foundations})\textbf{:} Williams--Beer, BROJA, Gaussian PID, Flow-PID}, fill=green!7, draw=green!40]
      [{\textbf{Interaction quantification} (\S\ref{sec:interaction-measurement})\textbf{:} CVX/BATCH, LSMI, ICYM2I, InterSHAP}, fill=green!7, draw=green!40]
      [{\textbf{PID disentanglement} (\S\ref{sec:interaction-representation})\textbf{:} MISA, FINE, DrFuse, PIBD, MRdIB}, fill=green!7, draw=green!40]
      [{\textbf{Interaction-typed MoE} (\S\ref{sec:interaction-moe})\textbf{:} MMOE, I2MoE, PathMoE}, fill=green!7, draw=green!40]
    ]
    [{Scaling to\\Foundation Models\\(\S\ref{sec:foundation})\\\mdseries\scriptsize capstone: P1--P3\\\mdseries\scriptsize on pretrained models}, fill=violet!16, draw=violet!55
      [{\textbf{Efficiency} (\S\ref{sec:foundation-efficiency})\textbf{:} MI-Pruner, InfoTok, MERGE, MINT}, fill=violet!7, draw=violet!35]
      [{\textbf{Interpretability} (\S\ref{sec:foundation-interpretability})\textbf{:} M2IB, NIB, DiffusionPID}, fill=violet!7, draw=violet!35]
      [{\textbf{Failure diagnosis} (\S\ref{sec:foundation-diagnosis})\textbf{:} GMI-Wasserstein bound, Vittle, VLV}, fill=violet!7, draw=violet!35]
    ]
  ]
\end{forest}
}
\caption{Taxonomy of the surveyed literature, read left to right as principle $\Rightarrow$ challenge $\Rightarrow$ method family $\Rightarrow$ representative methods. Each principle (P1--P3) motivates one challenge; the fourth challenge is a capstone applying all three to pretrained models. Branches are color-coded by challenge: alignment blue, robust fusion orange, interaction typing green, foundation models violet.}
\label{fig:taxonomy}
\end{figure*}
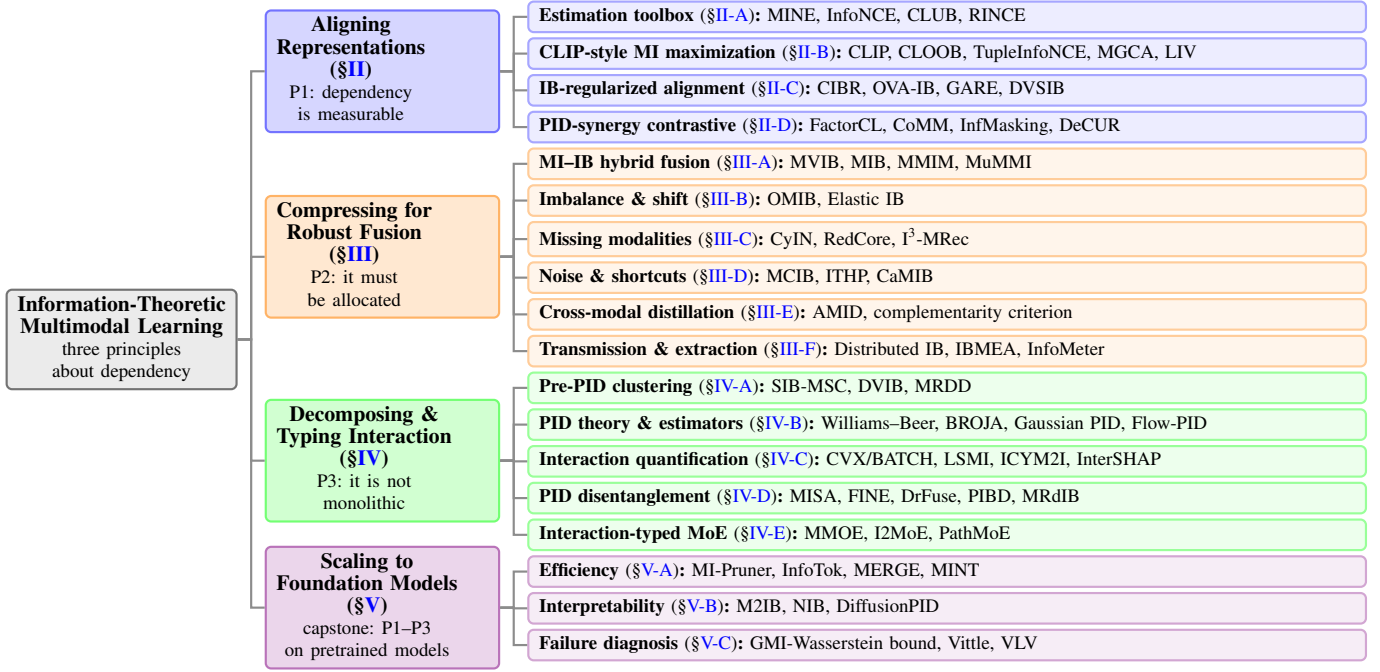

Information theory offers a task-agnostic vocabulary for all four. Mutual Information (MI) gives a single, estimator-agnostic quantity for how much two modalities, or a representation and a label, statistically depend on one another, independent of the encoder that produced the representation. The Information Bottleneck (IB) turns that quantity into a design principle, trading compression against prediction so a representation retains only what a task needs. Partial Information Decomposition (PID) splits a joint predictive dependency into redundant, unique, and synergistic parts. The three thus pose distinct questions: \emph{how much} two variables depend, \emph{how much} of that a representation should keep, and \emph{what kind} of dependency is kept. The survey's title names these three questions rather than the tools: MI, IB, and PID are the successive answers the field has given as its understanding of multimodal interaction has sharpened. Fig.~\ref{fig:taxonomy} gives the full taxonomy; Sections~\ref{sec:align}--\ref{sec:foundation} show this progression problem by problem.

The design principle governing Sections~\ref{sec:align}--\ref{sec:foundation} is this: \emph{a taxonomy should be derived from a small number of foundational principles about what the underlying quantities are and motivate, not from clustering the literature by keyword or application domain.} A domain name records where a method was validated, not the mechanism that makes it work; a taxonomy built on it conflates methods that solve different problems in the same domain and separates methods that solve the same problem in different domains. The principle has precedent -- \citet{Liang2024Foundations} derive their multimodal-ML taxonomy the same way, from what modalities are (heterogeneous, connected, interacting) -- but ours concern \emph{dependency between} modalities rather than modalities themselves, and the challenges they motivate are information-theoretic, not functional. Deriving a taxonomy from principles is nonetheless this survey's method, not its contribution -- the contribution is the unification itself. Table~\ref{tab:survey-comparison} positions this survey against the closest prior reviews. General multimodal surveys \citep{Baltrusaitis2019Multimodal,Liang2024Foundations,Yuan2025Survey} organize the field by functional task or modality properties and invoke information theory rarely or only in passing; framework-specific treatments each stay within a single tool -- variational MI bounds \citep{Poole2019Variational}, the Information Bottleneck \citep{Goldfeld2020Information}, or Partial Information Decomposition \citep{Lizier2018Information} -- and none is specific to multimodal learning. To our knowledge no survey treats MI, IB, and PID together as one progressive account of multimodal interaction. Section~\ref{sec:synthesis} makes that unification formal, defining a single objective -- the Generalized Multimodal Information Lagrangian (GMIL) -- in which the alignment, robust-fusion, and interaction-typing literatures occupy exact or approximate parameter corners. Two elementary structural observations locate where compression and interaction typing constrain each other. As a coordinate system over a broad class of linearly weighted bivariate typed-information objectives, the GMIL is not only descriptive: it localizes the parameter-space corners no surveyed method occupies and reads each as a concrete candidate method family (Table~\ref{tab:gmil-predicted}), and suggests a heuristic design guide mapping a task's measured interaction profile to candidate method families worth testing (Table~\ref{tab:decision}). That the same coordinate system both places the 2018--2026 corpus and charts the regions it leaves unoccupied is the survey's main contribution.

\begin{table}[t]
\centering
\caption{This survey against the most closely related prior reviews. \emph{MM}: specific to multimodal learning; MI/IB/PID: substantive coverage (\checkmark) vs.\ passing mention ($\circ$); \emph{U}: explicit formal account of how the frameworks relate.}
\label{tab:survey-comparison}
\setlength{\tabcolsep}{4pt}
\begin{tabularx}{\columnwidth}{@{}Xccccc@{}}
\toprule
Survey & MM & MI & IB & PID & U \\
\midrule
\citet{Baltrusaitis2019Multimodal} & \checkmark &  &  &  &  \\
\citet{Liang2024Foundations} & \checkmark & $\circ$ &  & $\circ$ &  \\
\citet{Yuan2025Survey} & \checkmark & $\circ$ & & & \\
\citet{Poole2019Variational} &  & \checkmark & $\circ$ &  &  \\
\citet{Goldfeld2020Information} &  & \checkmark & \checkmark &  &  \\
\citet{Lizier2018Information} & & & & \checkmark & \\
\midrule
This survey & \checkmark & \checkmark & \checkmark & \checkmark & \checkmark \\
\bottomrule
\end{tabularx}
\end{table}

Applying this to information-theoretic multimodal learning yields three principles about what dependency between modalities is, and four challenges they motivate. Principle~1, that dependency is measurable at all, motivates \emph{Aligning Representations} (Section~\ref{sec:align}): a measurable dependency is available as a training signal, and alignment is the challenge of exploiting it to bring representations into correspondence. Principle~2, that measured dependency must be allocated between what is discarded and retained, motivates \emph{Compressing for Robust Fusion} (Section~\ref{sec:robust}): a designer must decide how much survives compression, imbalance, noise, or missingness. Principle~3, that retained dependency is not monolithic -- it can be redundant across modalities, unique to one, or synergistic between them -- motivates \emph{Decomposing and Typing Interaction} (Section~\ref{sec:interaction}): a designer must know not just how much a fused representation retains but what kind, since ignoring modality-unique task-relevant information -- which much contrastive-learning work implicitly assumes away -- measurably degrades performance \citep{Liang2023Factorized}. A fourth, meta-level challenge, \emph{Scaling to Foundation Models} (Section~\ref{sec:foundation}), emerges once all three tools are mature: MI, IB, and PID are applied to representations already inside a pretrained model (as post-hoc probes or during lightweight adaptation, routing, or data selection), a capstone sitting above, not beside, the first three. These three principles order the four challenges into a conceptual progression rather than a list -- increasing resolution, not strict precedence, since PID can apply directly to raw modalities, as Section~\ref{sec:synthesis} does. Each challenge sharpens the quantity the previous one leaves coarse: alignment supplies the correspondence compression acts on, and compression leaves a scalar that interaction typing decomposes.

The three formal objects these principles invoke -- mutual information and its neural estimators, the Information Bottleneck Lagrangian, and the Partial Information Decomposition identity -- are defined where they are first used rather than in a standalone preliminaries section. Throughout, we treat encoders as stochastic unless stated otherwise: for deterministic encoders with non-atomic continuous outputs, $I(X;Z)$ can be infinite under common non-degeneracy conditions, so stochastic encoders are standard in variational IB methods, although deterministic variants also exist. We write $I(X;Y)$ for mutual information, which is symmetric, invariant under invertible reparameterization, and non-increasing under the data-processing inequality: the properties that make a bound on $I(X;Z)$ constrain any encoder producing $Z$ from $X$ regardless of architecture.

Consequently this survey has no separate ``applications by domain'' section. Domain-specific work -- medical imaging and report generation, robotics and embodied control, sentiment and emotion analysis, recommendation, knowledge-graph entity alignment, wireless semantic communication, and others -- appears throughout Sections~\ref{sec:align}--\ref{sec:foundation}, but strictly as evidence that a mechanism (an InfoNCE-style alignment loss, an IB compression term, a PID-typed mixture-of-experts router) generalizes beyond the benchmark it was first validated on: IB-regularized fusion for chest X-ray diagnosis and the identical mechanism for text-video retrieval sit side by side in Section~\ref{sec:robust}.

\textit{Scope and selection.} This survey covers 170 works published between 2018 and 2026 -- the majority discussed individually, the remainder cataloged by mechanism in each section's ``additional surveyed methods'' paragraph and listed in supplementary Table~S1 -- together with 12 earlier foundational references, from the classical Information Bottleneck \citep{Tishby1999Information} to the Partial Information Decomposition framework \citep{Williams2010Nonnegative}. The 2018--2026 span begins with the estimators that made neural mutual information tractable -- MINE \citep{Belghazi2018Mutual} and InfoNCE-based contrastive predictive coding \citep{Oord2018Representation} -- runs through the CLIP-era growth of contrastive alignment \citep{Radford2021Learning}, the multi-view Information Bottleneck \citep{Federici2020Learning}, and PID's operationalization for multimodal interaction analysis \citep{Liang2023Quantifying}, and ends with the 2024--2026 shift toward applying all three to pretrained multimodal large language models (MLLMs). We gathered papers by querying arXiv, the ACL Anthology, and major vision, machine-learning, and medical-imaging venues (the corpus spans CVPR/ICCV, NeurIPS/ICML/ICLR, ACL/EMNLP, and IEEE TMI/MICCAI, among others) with combinations of ``multimodal''/``multi-view'' and an explicit information-theoretic term -- mutual information, information bottleneck, partial information decomposition, redundancy, uniqueness, synergy, interaction information -- plus forward- and backward-citation tracing from the works above; the corpus covers papers appearing through mid-2026. Works available only as preprints at that cutoff are marked by their arXiv identifiers in the bibliography; empirical figures quoted from them are the papers' self-reported results. We include a paper if it makes an explicit information-theoretic quantity -- an entropy, divergence, mutual-information bound, or PID atom -- part of its formal objective, or explicitly operationalizes information-theoretic interaction types through a surrogate (as prediction-discrepancy or Shapley-interaction methods do), excluding work that uses information-theoretic language merely informally; the survey therefore omits the broader fusion literature relying on non-Shannon dependence measures such as kernel- or optimal-transport-based alignment, except where a surveyed method invokes them. Multimodal generative models whose objective is a likelihood-based ELBO rather than an explicit MI, IB, or PID term -- notably product- and mixture-of-experts multimodal VAEs \citep{Wu2018Multimodal,Shi2019Variational} -- sit adjacent to this scope and are noted where a surveyed method builds on them. Each paper is assigned to the single framework (MI, IB, or PID) its central objective instantiates; the few combining two (e.g., IB compression followed by PID-typed terms) are assigned by primary contribution and discussed under both where relevant (Table~\ref{tab:robust-methods} marks nominally-MI-but-IB-arriving methods MI$\to$IB).

The surveyed papers spread unevenly across sections and frameworks, yet no section is exclusive to one framework: even Decomposing and Typing Interaction (Section~\ref{sec:interaction}), where PID dominates, retains a pre-PID MI/IB cluster (\S\ref{sec:interaction-preformal}) that the rest of the section shows PID refining.

\section{Aligning Representations}
\label{sec:align}

Cross-modal alignment trains encoders so that paired observations of the same underlying content map to comparable points in representation space. MI maximization supplies the dominant training signal, IB principles repair what it leaves behind, and PID exposes an assumption both share (Table~\ref{tab:align-methods}).

\begin{table*}[t]
\centering
\footnotesize
\caption{Representative alignment methods of Section~\ref{sec:align}, spanning all four subsections. ``Frmwk.'': the framework (MI, IB, or PID) through which the method is primarily presented.}
\label{tab:align-methods}
\begin{tabularx}{\textwidth}{@{}m{2.5cm}@{\hspace{7pt}}>{\centering\arraybackslash}m{0.8cm}@{\hspace{7pt}}X@{\hspace{7pt}}m{4.25cm}@{\hspace{7pt}}>{\centering\arraybackslash}m{0.9cm}@{}}
\toprule
Method & Frmwk. & Mechanism & Modalities / domain & Subsec. \\
\midrule
MINE \citep{Belghazi2018Mutual} & MI & Donsker--Varadhan MI lower bound, trained as neural statistics network & General (continuous representations) & \S\ref{sec:align-toolbox} \\
InfoNCE / CPC \citep{Oord2018Representation} & MI & Categorical cross-entropy contrastive lower bound (1 positive vs.\ $N{-}1$) & General (default alignment objective) & \S\ref{sec:align-toolbox} \\
CLUB \citep{Cheng2020CLUB} & MI & Contrastive log-ratio MI \emph{upper} bound (for compression / unique information) & General (compression/PID estimator) & \S\ref{sec:align-toolbox} \\
\midrule
CLIP \citep{Radford2021Learning} & MI & Symmetric InfoNCE alignment, no compression term & Image--text (400M web pairs) & \S\ref{sec:align-clip} \\
OVA-IB \citep{Li2026OVA} & IB & One-vs-All InfoNCE (Dual Total Correlation) + upper-bound minimality regularizer & Arbitrarily many modalities & \S\ref{sec:align-ib} \\
\midrule
FactorCL \citep{Liang2023Factorized} & PID & Shared/unique factorization via InfoNCE lower + NCE-CLUB upper bounds & General multimodal & \S\ref{sec:align-pid} \\
CoMM \citep{Dufumier2025What} & PID & InfoNCE between augmented fused views; preserves total $R{+}U_1{+}U_2{+}S$ (under stated assumptions) & General multimodal (Trifeature) & \S\ref{sec:align-pid} \\
\bottomrule
\end{tabularx}
\end{table*}

\subsection{The Shared Estimation Toolbox}
\label{sec:align-toolbox}

Mutual information $I(X;Y)=H(X)-H(X\mid Y)$, equivalently the KL divergence between the joint and the product of marginals \citep{Shannon1948Mathematical,Cover2006Elements}, is the architecture-independent dependency measure Principle~1 posits; it is generally difficult to estimate in high-dimensional neural settings, for which variational bounds provide a common scalable approach \citep{Poole2019Variational}. \citet{Belghazi2018Mutual}'s MINE lower-bounds MI via the Donsker--Varadhan dual of KL divergence, establishing the template of MI-estimation-as-trainable-objective. \citet{Oord2018Representation}'s InfoNCE, a density-ratio classifier distinguishing one positive from $N-1$ negatives, provably lower-bounds MI and, for its stability, became the default estimator across \S\ref{sec:align-clip} and~\ref{sec:align-ib}.

Because they are lower bounds, minimizing InfoNCE or MINE does not in general guarantee a reduction in the true mutual information, which is what compression requires. \citet{Cheng2020CLUB} address this limitation with the Contrastive Log-ratio Upper Bound,
\[
\begin{aligned}
I_{\mathrm{CLUB}}(x;y) = {}&\mathbb{E}_{p(x,y)}[\log p(y|x)]\\
&- \mathbb{E}_{p(x)}\mathbb{E}_{p(y)}[\log p(y|x)],
\end{aligned}
\]
provably an upper bound, on which later IB-style compression and PID-style unique-information bounds build. \citet{Chuang2022Robust} add RINCE, a tunable loss family lower-bounding the Wasserstein Dependency Measure that interpolates toward a symmetric, noise-robust loss for noisy correspondences that violate InfoNCE's shared-information assumption.

\subsection{CLIP-Style Contrastive Alignment: Pure MI Maximization}
\label{sec:align-clip}

The canonical pure-MI case is CLIP \citep{Radford2021Learning}, training dual image and text encoders on 400 million web pairs with a symmetric InfoNCE objective over the $N$ correct and $N^2-N$ incorrect pairings in a batch. Its zero-shot transfer -- matching a supervised ResNet-50 on ImageNet without labels -- established the paradigm the rest of this subsection extends, none adding explicit compression.

Several works repair InfoNCE failure modes: \citet{Frst2022CLOOB}'s CLOOB swaps the saturating alignment/uniformity ratio for the non-saturating InfoLOOB bound; \citet{Chen2023ProtoCLIP} add a prototype-level discrimination loss; \citet{Qin2021TVDIM} substitute a margin-based ranking loss tolerant of imperfect pairs. Two further variants target robustness: \citet{Ma2021Active} mine high-uncertainty negatives to ease the sample-complexity limits of MI lower bounds, and \citet{Rahman2025DiMPLe} add a conditional-MI-minimization step (via HSIC) to separate invariant from spurious features. A second cluster goes beyond two variables: \citet{Liu2021Contrastive} extend InfoNCE to $K$-modality tuples (TupleInfoNCE), and \citet{Koutoupis2025More} add a fused-modality total-correlation term that solves synthetic synergy tasks pairwise CLIP fails at. Both maximize a total-correlation quantity rather than decomposing it into atoms, setting up \S\ref{sec:align-pid}.

Three papers characterize CLIP's optimum. \citet{Gui2025Multi} characterize global temperature-optimized InfoNCE minimizers as having intrinsic dimension equal to that of the ideal shared representation (under population-level realizability assumptions: a sufficiently rich function class and the existence of aligned maximal-information representations), which explains why effective dimensionality saturates below the embedding size and temperature converges to zero. \citet{Cai2026Geometric} show that symmetric InfoNCE induces a repulsive symmetric-KL coupling between the modality marginals, a geometric account of the ``modality gap'' (derived in the large-batch population-energy limit under compactness and regularity assumptions). \citet{Uesaka2025Weighted} prove the optimal similarity function is exactly the pointwise mutual information (PMI), which is not exactly representable by a bilinear similarity of $d$-dimensional features once the target PMI matrix over $N$ pairs has rank exceeding $d+1$. And as \S\ref{sec:align-ib} shows formally, MI maximization provides no guarantee of removing modality-specific nuisance.

The same recipe recurs wherever an application supplies two aligned streams. \citet{Wang2022Multi}'s MGCA jointly optimizes instance-, token-, and disease-prototype-level InfoNCE objectives over chest X-rays and radiology reports, and \citet{Ma2023LIV}'s LIV combines goal-conditioned value learning with CLIP's InfoNCE loss, proving the value loss reduces exactly to CLIP's when training videos collapse to a single caption-aligned goal frame. Across these and further deployments the loss stays an identical InfoNCE-family objective; only corpus and downstream task vary.

\subsection{Adding Minimality: IB-Regularized Alignment}
\label{sec:align-ib}

Under a multi-view-redundancy or equivalent task-alignment assumption, pure cross-modal MI maximization can support task sufficiency but does not enforce minimality, retaining modality-specific nuisance InfoNCE has no incentive to discard. The papers here add an IB-style compression term atop the MI objective.

\citet{Almudvar2025Aligning} formalize each modality as shared ``essence'' plus modality-specific ``nuisance,'' prove InfoNCE does nothing to enforce $I(Z_\alpha;N_\alpha)=0$ (the mechanism behind the modality gap), and derive, for modalities $\alpha\neq\beta$, an IB objective $\max I(Z_\alpha;X_\beta) - \lambda I(Z_\alpha;X_\alpha)$ with a variational KL upper bound on its intractable second term. \citet{Ji2025CIBR} argue, through a proposition and proof sketch, that CLIP approximately performs an implicit cross-modal bottleneck, and add an explicit MINE-estimated penalty,
\[
L_{\mathrm{CIBR}} = L_{\mathrm{CLIP}} + \lambda\big[I(Z_v;X_v\mid X_t) + I(Z_t;X_t\mid X_v)\big],
\]
so minimizing $L_{\mathrm{CIBR}}$ drives down residual modality-specific information, with consistent gains over CLIP, CoOp, and MaPLe. \citet{Li2026OVA}'s OVA-IB generalizes to arbitrarily many modalities, beating summed pairwise CLIP losses by margins that widen with modality count.

\citet{Xiao2025Rebalancing}'s GARE diagnoses a gradient-level pathology in text-video retrieval, where positive and negative InfoNCE gradients cancel, and adds a pair-specific correction regularized as a deterministic variational IB. \citet{Dang2024Disentangled} decompose MI into modality-invariant and modality-exclusive parts for noisy correspondence, reducing at one setting exactly to a multi-view IB objective. Two papers unify the family: \citet{Wen2024MVEB}'s MVEB recasts the multi-view IB Lagrangian as a tractable alignment-plus-differential-entropy objective subsuming InfoNCE, BYOL, and Barlow Twins \citep{Zbontar2021Barlow}, while \citet{Abdelaleem2023Deep}'s DVSIB casts VAEs, CLIP, and Barlow Twins as one $\mathcal{L}=I_{\text{encoder}}-\beta I_{\text{decoder}}$ family, provably reducing to both CLIP and Barlow Twins under deterministic assumptions. CLIP's objective is thus a special case of a broader IB-regularized family, and the surveyed results show that explicit compression can improve alignment models in several task- and architecture-specific settings.

\subsection{Beyond Redundancy: PID-Informed Synergistic Contrastive Learning}
\label{sec:align-pid}

\S\ref{sec:align-clip} and~\ref{sec:align-ib} disagree about compression but share an unstated assumption: everything relevant downstream is redundantly present in both modalities, so cross-modal MI is the right target. \citet{Liang2023Factorized} name this the ``multi-view redundancy assumption'' and prove, via a Bayes-error bound, that contrastive learning degrades as modality-unique task-relevant information grows. The papers here break the assumption using PID's decomposition of $I(X_1,X_2;Y)$ into redundant ($R$), unique ($U_1,U_2$), and synergistic ($S$) components (Section~\ref{sec:interaction}).

FactorCL \citep{Liang2023Factorized} factorizes representations into task-relevant shared and unique components \citep{Cheng2020CLUB} but neither targets synergy explicitly nor escapes an impractical optimal-augmentation assumption. \citet{Dufumier2025What}'s CoMM theoretically preserves the total task-relevant information $R+U_1+U_2+S$ (assuming minimal label-preserving augmentations, sufficient encoder expressivity, and optimization to the MI optimum). It does not identify the four atoms separately, yet alone captures all three interaction types in controlled Trifeature experiments. Arguing CoMM under-captures synergy, \citet{Wen2025InfMasking}'s InfMasking stochastically masks large proportions of each modality before fusion, raising synergy-task accuracy from 71.4\% to 77.0\% on a controlled synergy benchmark, while \citet{Cisse2026Orthogonalized} add a shared-unique orthogonality constraint plus asymmetric masking for the best uniqueness recovery here. \citet{Song2024QUEST} form a quaternion cross-product embedding of the components, and \citet{Kontras2026SynIB}'s SynIB instead \emph{induces} synergy by penalizing confident predictions under modality-masking counterfactuals.

Two papers come from self-supervised redundancy reduction. \citet{Mohamadi2023More} reframe two-view SSL as a three-variable MI problem and show, via a Williams-and-Beer PID decomposition \citep{Williams2010Nonnegative}, that Barlow Twins' \citep{Zbontar2021Barlow} whitening losses destroy synergistic information along with redundant, recovering it with a recalibration protocol. \citet{Wang2024Decoupling}'s DeCUR instead partitions Barlow Twins' embedding space directly into common and unique subspaces, validated on SAR-optical and RGB-depth data.

\emph{Additional surveyed methods.} Beyond those above, the literature includes further MI-estimation refinements and benchmarks \citep{Hashmani2025Multimodal,Hjelm2019Learning,Barber2003IM,Song2020Understanding} and many domain deployments of the same InfoNCE-family mechanism: medical image--text pretraining and cross-protocol registration \citep{Li2024MLIP,Snaauw2022Mutual,Wang2023Zero}, composed retrieval and cross-modal hashing \citep{Gu2022Cross,Hoang2023Multi}, audio-visual segmentation \citep{Mao2023Contrastive}, embodied skill discovery and wearable activity recognition \citep{Ju2024Rethinking,Matsuishi2025Multimodal}, and generation evaluation \citep{Kim2022Mutual}.

\section{Compressing for Robust Fusion}
\label{sec:robust}

Naive MI-maximizing fusion can become brittle when modalities are missing, corrupted, or strongly imbalanced -- when sensors fail, one modality starves its partner of gradient, or annotation artifacts create shortcuts. This section surveys work confronting this brittleness (Table~\ref{tab:robust-methods}), grouped by the sub-problem each compression mechanism targets. A recurring finding: ``mutual information'' and ``information bottleneck'' work often solve identical problems with near-identical machinery, differing mainly in whether the compression term is an explicit Lagrangian or folded into an architecture, regularizer, or distillation loss.

\begin{table*}[t]
\centering
\footnotesize
\caption{Representative robust-fusion methods of Section~\ref{sec:robust}, spanning all six subsections. ``Frmwk.'': MI, IB, or MI$\to$IB (nominally MI-based but arriving at an IB-style design).}
\label{tab:robust-methods}
\setlength{\tabcolsep}{4pt}
\begin{tabularx}{\textwidth}{@{}m{2.2cm}>{\centering\arraybackslash}m{0.9cm}m{1.8cm}X>{\centering\arraybackslash}m{1.2cm}@{}}
\toprule
Method & Frmwk. & Sub-problem & Compression mechanism & Subsec. \\
\midrule
MVIB \citep{Federici2020Learning} & MI$\to$IB & hybrid & MI lower bound + symmetrized-KL penalty on view-private information (multi-view IB) & \S\ref{sec:robust-hybrid} \\
MIB \citep{Mai2023Multimodal} & IB & hybrid & IB Lagrangian $I(y;z){-}\beta I(x;z)$ ($\beta$ on compression) over unimodal / fused embeddings & \S\ref{sec:robust-hybrid} \\
\midrule
OMIB \citep{Wu2025Learning} & IB & imbalance & Upper bound on $\beta$ under a disjoint information-set assumption; per-modality weight from conditional task information & \S\ref{sec:robust-imbalance} \\
\midrule
CyIN \citep{Lin2025CyIN} & IB & missing & Cyclic variational-IB purification + cyclic-consistency latent reconstruction & \S\ref{sec:robust-missing} \\
\midrule
MCIB \citep{Wang2026Conditional} & IB & noise-shortcut & Conditional IB (Wyner CMI): compress primary, keep only complementary information & \S\ref{sec:robust-noise} \\
\midrule
\citet{Xie2025Information} & MI & distillation & Complementarity criterion: distillation helps if $I(H_1;H_2) > I(H_2;Y)$ (sufficient condition; Gaussian asymptotic model) & \S\ref{sec:robust-distillation} \\
\midrule
\citet{Zhou2025Distributed} & IB & transmission & Distributed IB + learnable per-task modality-task-link pruning operator & \S\ref{sec:robust-transmission} \\
\bottomrule
\end{tabularx}
\end{table*}

\subsection{Foundational MI--IB Hybrid Fusion Architectures}
\label{sec:robust-hybrid}

\citet{Tishby1999Information}'s Information Bottleneck seeks a stochastic $Z$ minimizing $I(X;Z)-\beta\,I(Z;Y)$, $\beta$ tuning the compression--relevance trade-off; its variational form \citep{Alemi2017Deep} made it usable in deep encoders. (Whether IB \emph{explains} deep-network generalization remains debated \citep{Tishby2015Deep,Saxe2019Information}; the methods here use it as a design principle, not a theory of learning.) Several of the most influential papers here, though nominally MI-framework, reach an IB-style design -- an MI-maximizing signal plus a term discarding what it does not need -- while others stop at pure maximization. \citet{Federici2020Learning}'s multi-view IB (MVIB) set the template, maximizing an MI lower bound between two views while penalizing their symmetrized KL divergence to discard view-private information. \citet{Mai2023Multimodal}'s MIB makes this explicit for sentiment and emotion recognition via the maximized IB Lagrangian $I(y;z) - \beta I(x;z)$ ($\beta$ here weighting compression), with explicit compression beating uncompressed tensor or graph fusion. \citet{Han2021Improving} supply the widely extended MMIM baseline maximizing MI between unimodal pairs and between the fused representation and each input.

The recipe re-emerges far from sentiment analysis. In reinforcement learning, MuMMI \citep{Chen2021Multi} replaces a world model's reconstruction terms with an InfoNCE objective pulling each sensor expert toward a shared fused code; \citet{You2024Multimodal} add the compression half MuMMI lacks, bounding MI between compressed state and raw inputs via Deep-VIB. \citet{Tian2023Variational} sidestep estimator instability, recasting the multi-view IB Lagrangian into purely KL-divergence terms that scale to arbitrarily many views without an MI estimator, whereas \citet{Yu2026Multimodal} generalize MINE's Donsker--Varadhan bound to total correlation among all modalities and the label with no compression term -- not every fusion-balancing method needs an explicit bottleneck.

\subsection{Correcting Modality Imbalance and Distribution Shift}
\label{sec:robust-imbalance}

A second failure mode arises even with all modalities present and clean: one converges faster or is more discriminative, dominates joint training, and suppresses the weaker. \citet{Gao2025Asymmetric} detect this with a lower-bound MI valuation of each modality's asymmetric contribution, driving a balanced min-max loss that reinforces weak modalities while empirically mitigating the ``dominant-modality forgetting'' of weak-side-only methods.

The IB response, OMIB \citep{Wu2025Learning}, reframes the problem from first principles. Where prior multimodal IB methods -- including \citet{Mai2023Multimodal}'s -- set $\beta$ ad hoc and assume modality symmetry, OMIB partitions multimodal information into consistent, modality-specific, and superfluous components and, within its idealized disjoint information-set assumption, proposes an upper bound on $\beta$ intended to make a minimal-sufficient bottleneck achievable (its appendix derivation switches the relevance term from MI to a normalized relative-MI objective). It then replaces fixed $\beta$ with a per-modality weight from each modality's conditional task-relevant information, setting the weight analytically rather than correcting imbalance reactively.

On \emph{distribution shift}, \citet{Ni2022Elastic} decompose target-domain error into training error, a generalization gap, and a representation-discrepancy term, showing the deterministic IB variant (compressing via $H(T)$ rather than $I(X;T)$) yields a smaller generalization gap while stochastic IB yields smaller cross-domain discrepancy; their Elastic IB interpolates via a mixing coefficient $\alpha$, and on MNIST-to-USPS transfer intermediate values dominate either pure objective.

\subsection{Robustness to Missing Modalities}
\label{sec:robust-missing}

A distinct failure mode arises when a modality is entirely absent at inference; nearly every paper here is IB-framework, reconstructing missing information in IB-purified feature space. \citet{Lin2025CyIN}'s CyIN builds one unified model rather than separate per-missing-pattern models, applying the variational IB bound cyclically and reconstructing a missing modality's bottleneck latent from the remaining purified latents. \citet{Sun2024RedCore} isolate a subtler variant: under different missing rates, uniform imputation under-supervises the modality missing more often, so RedCore sets each modality's supervision by its ``relative advantage'' via a bi-level optimization whose supervision-weight subproblem is equivalent to a convex problem for fixed model parameters. \citet{Chen2025MRec}'s I$^3$-MRec confronts missing item images or descriptions in recommendation, combining invariant risk minimization with a missing-aware IB fusion term and degrading only $\sim$12\% in Recall@20 at 90\% missing.

\subsection{Robustness to Noise, Shortcuts, and Unreliable Modalities}
\label{sec:robust-noise}

A subtler failure mode than outright absence: a modality can be present yet corrupted by noise or correlated with the label only through a spurious shortcut. \citet{Wang2026Conditional} document this for sarcasm detection, where MUStARD++-trained models exploit canned-laughter cues, character-role heuristics, and emotion-polarity leakage; their Multimodal Conditional Information Bottleneck (MCIB), built on Wyner's conditional mutual information \citep{Wyner1978Definition}, compresses each primary modality while retaining only what is complementary given the auxiliary modality, verified on a de-shortcutted benchmark (MUStARD++$^R$). \citet{Xiao2024Neuro}'s ITHP instead targets redundancy, cascading IB bottlenecks across modalities as sequential ``detectors.''

The most pointed critique comes from CaMIB \citep{Jiang2025Minimal}: the dominant ``learning-to-attend'' paradigm conflates genuine cross-modal synergy with statistical shortcuts, since both increase the same MI objective. CaMIB formalizes shortcut information as an unobserved confounder in a structural causal model and combines per-modality IB filtering with a self-attention-derived instrumental variable and backdoor adjustment to break the causal-shortcut correlation, improving out-of-distribution accuracy. An MI-only objective, blind to what is discarded, is the shared vulnerability across \S\ref{sec:robust-hybrid}--\S\ref{sec:robust-noise}; CaMIB adds that it also cannot separate causal predictive information from dataset artifact.

\subsection{Compression as Cross-Modal Knowledge Distillation}
\label{sec:robust-distillation}

Another failure mode is an imperfect deployment budget: only a single, weaker modality may be available at inference despite richer training data. Cross-modal distillation trains a teacher on the rich set and distills into a weak-modality student, reframed as maximizing the MI the student retains about the teacher without collapsing into a shortcut. \citet{Chen2023Enhanced} diagnose exactly such a collapse -- maximizing $I(\text{teacher};\text{student})$ alone lets the teacher degrade toward the student's limited informativeness -- and their AMID adds a second MI term to an auxiliary-modality model plus an adversarial entropy bound to prevent it. \citet{Xie2025Information} give a testable complementarity criterion: distillation helps when teacher-student MI $I(H_1;H_2)$ exceeds student-label MI $I(H_2;Y)$ -- a sufficient condition proved via an excess-risk bound in a jointly Gaussian, small-$\lambda$ asymptotic model, with the converse observed only empirically. Injecting noise to flip the gap's sign eliminates the benefit.

\subsection{Compression for Efficient Transmission, Extraction, and Retrieval-Adjacent Tasks}
\label{sec:robust-transmission}

The most heterogeneous cluster applies the same compress-what-is-not-predictive logic to tasks sharing a resource constraint -- bandwidth, label budget, or interpretability -- rather than a failure mode. In wireless semantic communication, \citet{Zhou2025Distributed} extend the Distributed Information Bottleneck (Distributed IB) with a learnable modality-selection operator that prunes low-relevance modality-task links, cutting active links by up to 70\% and communication rate by over 40\% while matching full-participation baselines. \citet{Tian2021Farewell} argue unstable high-dimensional MI estimation is itself the weak link in person re-identification, recasting the IB sufficiency objective as an equivalent KL divergence -- fitting MI without estimating it. \citet{Su2024IBMEA}'s IBMEA applies per-modality variational IB to knowledge-graph entity alignment, suppressing alignment-irrelevant ``misleading clues'' attention-based fusion cannot discard. Finally, \citet{Hadizadeh2024Mutual}'s InfoMeter finds that \emph{lower} camera-LiDAR MI -- greater complementarity than redundancy -- consistently predicts higher 3D-detection accuracy in their experiments, providing suggestive support for the redundancy-suppressing mechanisms surveyed here.

Whether or not a method cites Tishby's Information Bottleneck \citep{Tishby1999Information}, most objectives here remain scalar or untyped at the level of each information term -- which Section~\ref{sec:interaction} takes apart into redundant, unique, and synergistic parts.

\emph{Additional surveyed methods.} Further robust-fusion work spans hybrid MI--IB architectures \citep{Wu2023Denoising,Jiang2023Correlation,Fu2025Robust,Kontras2025Multimodal,Naderi2026Joint,Becker2024Combining,Nguyen2023Self,Zheng2022Multimodal,Shen2024Complementary,Fang2024Dynamic,Nguyen2010Estimating}, imbalance and distribution-shift correction \citep{Xie2025Balanced,Sun2026Boomda,Zhang2026Semantic,Kieu2025Enhancing}, missing-modality robustness \citep{Xu2025Contribution,Li2025Theory,Li2024Robust}, noise and reliability-weighted fusion \citep{Liu2024RNG,Huang2025Robust,Xie2024Trustworthy}, cross-modal distillation \citep{Chen2021Distilling}, and transmission-, extraction-, and retrieval-adjacent applications \citep{Wei2025Task,Li2025OTCR,Zhou2025Multimodal,Shi2023Learning,Li2026Information,Kim2023Heterogeneous,Songara2023Visual,Ma2019Unpaired,Li2025NOFT}.

\section{Decomposing and Typing Interaction}
\label{sec:interaction}

When two modalities jointly predict a target, is the label recoverable from either alone (redundancy), does it require one modality's exclusive access to some fact (uniqueness), or does it emerge only from their joint observation (synergy)? For a decade the multimodal and multi-view literatures answered this only informally, using MI maximization or the IB to carve a representation into ``shared'' and ``private'' parts, with no vocabulary for ``private'' beyond ``not shared.'' PID supplies that vocabulary, splitting the joint predictive information into four non-negative terms via the identity $I(X_1,X_2;Y)=R+U_1+U_2+S$ -- redundancy, unique information, and synergy -- each with an operational meaning no single MI or IB scalar can express. PID-based methods reach beyond the sentiment benchmarks where they first took root, into medicine, recommendation, and federated learning.

\subsection{MI/IB-Based Multi-View Clustering and Disentanglement: A Pre-PID Approach}
\label{sec:interaction-preformal}

The multi-view clustering literature was arguably first to confront this at scale: separate what several views of one entity share from what each carries alone, so the shared part drives consensus clustering while the private part is neither discarded nor injected as noise. Nearly every method reaches for the same two tools -- MI maximization to pull shared representations together, IB-style compression to strip away what is not shared.

\citet{Wang2022Self}'s SIB-MSC, the first purely IB-based multi-view subspace clustering method, splits one view's MI with the input into a superfluous component (bounded away) and a signal-relevant one (maximized), with a predictability-minimization branch separating view-specific from shared codes; the split only decides which MI term is maximized versus bounded, with no guarantee whether the shared code is redundant or synergistic. \citet{Bao2021Disentangled}'s Disentangled Variational Information Bottleneck (DVIB) is the clearest pre-PID statement of the paradigm -- four MI-based objectives make each shared latent informative about both views and each private latent informative only about its own view -- yet the privateness of each private latent is expressible only as a difference of MI bounds, without proven non-negativity. \citet{Ke2024Rethinking}'s two-stage MRDD targets the resulting leak between consistency and specificity, first learning a consistent encoder via masked cross-view prediction, then freezing it and training view-specific encoders that minimize a CLUB estimate of their MI with the consistent features.

The MI/IB toolkit measures amounts, not kinds; certifying whether ``shared'' information is redundant or genuinely synergistic is what PID was invented for.

\subsection{Foundational PID Theory and Estimators}
\label{sec:interaction-foundations}

The theoretical starting point is \citet{Williams2010Nonnegative}'s nonnegative decomposition of multivariate information. Classical measures such as the interaction information of \citet{McGill1954Multivariate} go negative once three or more variables are involved, conflating synergy with redundancy; Williams and Beer resolve this with a redundancy measure $I_{\min}$ inducing a redundancy lattice whose M\"obius inversion yields Partial Information atoms that non-negatively decompose the target--source mutual information into redundant, unique, and synergistic pieces. The atoms depend on the chosen redundancy measure: $I_{\min}$, the intersection-information measure of \citet{Griffith2015Quantifying}, and the operational unique-information measure of \citet{Bertschinger2014Quantifying} can disagree numerically, and non-negativity holds only for measures constructed to ensure it. Formally, for two sources and a target the decomposition is pinned down by three consistency equations in four unknowns -- $I(X_1,X_2;Y)=R+U_1+U_2+S$, $I(X_1;Y)=R+U_1$, $I(X_2;Y)=R+U_2$ -- closed by the choice of redundancy measure; Williams--Beer take $R=\sum_y p(y)\,\min_i I_{\mathrm{spec}}(y;X_i)$, where $I_{\mathrm{spec}}(y;X_i)$ is the specific information $X_i$ carries about outcome $y$. Two canonical cases fix intuition: for $Y=X_1=X_2$ a fair bit, $R=1$ and $U_1=U_2=S=0$; for $Y=X_1\oplus X_2$ with independent fair inputs, $R=U_1=U_2=0$ and $S=1$ -- each source alone is useless, the pair determines $Y$.

Subsequent work makes this decomposition practically estimable. The Bertschinger--Rauh--Olbrich--Jost--Ay (BROJA) unique-information measure \citep{Bertschinger2014Quantifying} was originally tractable only for discrete or Gaussian variables. \citet{Pakman2021Estimating} lift this via copula-parametrized variational upper bounds, extending PID estimation to arbitrary continuous distributions for the first time and revealing, on a generalized-XOR task, a trade-off between synergy-dominated and uniqueness-dominated (``grandmother cell'') codes. \citet{Venkatesh2023Gaussian} instead restrict BROJA to the jointly Gaussian case, collapsing the exponential search to a quadratic-in-dimension optimization with closed-form objective and analytic gradient while debiasing the four atoms. \citet{Zhao2025Partial} prove the BROJA optimum is exactly jointly Gaussian whenever the pairwise marginals are, recast the objective as a lower-complexity log-determinant optimization (Thin-PID), and extend it beyond Gaussian data with MI-preserving normalizing-flow encoders (Flow-PID).

A complementary strand asks what PID guarantees when it \emph{cannot} be computed exactly. \citet{Liang2024Multimodal} show that with only labeled unimodal plus unlabeled multimodal co-occurrence data, synergy is in general unrecoverable, and derive provable lower and upper bounds on it that accurately predict trained multimodal model performance without any model training. Together these results turn the informal ``shared vs.\ private'' intuition into a rigorous, four-term, non-negative decomposition ready for use as a measurement and design tool, the subject of the next three subsections.

\subsection{Frameworks for Quantifying Multimodal Interactions}
\label{sec:interaction-measurement}

The paper that ports PID into mainstream multimodal ML, treated as the field's hub, is \citet{Liang2023Quantifying}. Adopting the \citet{Bertschinger2014Quantifying} definition, it introduces two scalable estimators -- CVX (a convex KL minimization on discretized features) and BATCH (neural encoders with a differentiable Sinkhorn--Knopp projection for high-dimensional continuous data) -- showing PID-based model selection reaches 95--100\% of best-model performance across five synthetic and six real MultiBench datasets without training the candidates. \citet{Liang2023Multimodal} extend the estimator to a human-in-the-loop setting, converting partial or counterfactual human labels into $R/U/S$ estimates that agree with directly annotated decompositions.

Later papers extend this hub. \citet{Yang2025Efficient}'s LSMI is a pointwise, per-sample estimator that matches CVX/BATCH orders of magnitude faster, enabling sample-level uses a dataset-level estimate cannot support. \citet{Choi2026ICYM} expose a bias in PID computed on modality-complete data under realistic missing-at-random shifts between training and deployment; their ICYM2I framework applies inverse-probability weighting to both training/evaluation and PID estimation, showing on a clinical case study that naive analysis overstates a chest X-ray's unique diagnostic information nearly threefold (5\% versus 1.8\%). \citet{Wenderoth2025Measuring}'s InterSHAP instead replaces the PID lattice with a Shapley Interaction Index over modality coalitions, generalizing to any number of modalities and to individualized, performance-agnostic explanations; on a synthetic pure-synergy dataset it measures markedly more synergy than PID.

A scalable $R/U_1/U_2/S$ framework also serves as a measurement instrument far from the sentiment benchmarks (CMU-MOSI, CMU-MOSEI, MUStARD, UR-FUNNY) on which PID was first validated. \citet{Chen2026SynGR} apply the lattice to generative recommendation, targeting the synergy gap $I(X_v,X_t;Y) - [I(X_v;Y)+I(X_t;Y)-I(X_v;X_t)]$ -- a quantity that equals the conditional mutual information $I(X_v;X_t\mid Y)$, and is thus a synergy-linked proxy also reflecting redundancy and inter-modal dependence rather than the synergy atom itself; SynGR pursues it via saliency-aware masking of the dominant modality plus a synergistic InfoNCE objective, gaining roughly 10\% on average over alignment-centric recommenders.

\subsection{PID-Guided Representation Disentanglement}
\label{sec:interaction-representation}

Extending \S\ref{sec:interaction-preformal}'s shared/private goal, this subsection folds the typing built in \S\ref{sec:interaction-measurement} directly into representation-learning objectives (Table~\ref{tab:pid-methods}).

\begin{table*}[t]
\centering
\footnotesize
\caption{PID-guided methods operationalizing the redundancy/uniqueness/synergy taxonomy (\S\ref{sec:interaction-representation}--\S\ref{sec:interaction-moe}). The atoms column lists the components each method explicitly targets; the estimator column names the mechanism actually used, often a heuristic or distributional proxy rather than an exact PID estimator.}
\label{tab:pid-methods}
\begin{tabularx}{\textwidth}{@{}m{2.9cm}@{\hspace{7pt}}>{\centering\arraybackslash}m{1.1cm}@{\hspace{7pt}}X@{\hspace{7pt}}m{4.25cm}@{\hspace{7pt}}>{\centering\arraybackslash}m{0.9cm}@{}}
\toprule
Method & Atoms & Estimator / surrogate & Domain & Subsec. \\
\midrule
DisentangledSSL \citep{Wang2025Information} & $R$, $U$ & Constrained IB-style objective (optimal when MNI attainable; near-optimal under further conditions otherwise) & Self-supervised multimodal & \S\ref{sec:interaction-representation} \\
DrFuse \citep{Yao2024DrFuse} & $R$, $U$ & Jensen--Shannon-divergence alignment + orthogonality (MISA-style) & Clinical (EHR + chest X-ray) & \S\ref{sec:interaction-representation} \\
MRdIB \citep{Wang2025Multimodal} & $R$, $U$, $S$ & Multimodal IB + MINE-style redundancy discriminator + joint predictor & Multimodal recommendation & \S\ref{sec:interaction-representation} \\
Robult \citep{Nguyen2025Robult} & $R$, $U$ & Bivariate PID identity: soft PU-contrastive ($R$) + latent reconstruction ($U$) & Missing modality, label scarcity & \S\ref{sec:interaction-representation} \\
\midrule
MMOE \citep{Yu2024MMOE} & $R$, $U$, $S$ & Prediction-discrepancy heuristic (unimodal vs.\ multimodal label) & Affective (sarcasm) & \S\ref{sec:interaction-moe} \\
I2MoE \citep{Xin2025MoE} & $R$, $U$, $S$ & Weakly-supervised interaction loss (single-modality perturbation) + re-weighting & General multimodal & \S\ref{sec:interaction-moe} \\
PathMoE \citep{Yu2026PathMoE} & $R$, $U$, $S$ & I2MoE-style interaction loss with per-sample gating (5 experts) & Pathology (tumor subtyping) & \S\ref{sec:interaction-moe} \\
\bottomrule
\end{tabularx}
\end{table*}
 The bridge between the two eras is \citet{Hazarika2020MISA}, whose Modality-Invariant and -Specific representations (MISA) predate formal PID yet already project each modality into an invariant subspace (aligned via Central Moment Discrepancy) and a private subspace (kept non-redundant via soft orthogonality) before fusion. Like \S\ref{sec:interaction-preformal}'s methods it uses distributional-distance surrogates rather than a formal $R/U/S$ decomposition, yet became the canonical baseline nearly every PID-guided method below extends.

Several sentiment papers re-engineer MISA's mechanism. \citet{Qian2026DecAlign} replace MISA's global losses with a hierarchical scheme -- modality-unique features aligned via Gaussian-mixture-prototype optimal transport, modality-common features via moment-matching and MMD. \citet{Liu2026FINE}'s FINE isolates the unique branch with a paired InfoNCE lower bound and a CLUB upper bound, beating MISA and later contrastive baselines on CMU-MOSI, CMU-MOSEI, UR-FUNNY, and CH-SIMS. A second family adds an explicit optimality guarantee. \citet{Wang2025Information}'s DisentangledSSL derives shared and modality-specific representations via a constrained IB-style objective; these are exactly optimal when the Minimum Necessary Information point (a representation retaining exactly the shared task-relevant information and no more) is attainable, with bounded near-optimality guarantees under additional information-curve conditions when it is not. \citet{Ma2025Explainable}'s PIDReg enforces approximate joint Gaussianity of the latents to make the closed-form Gaussian PID of \S\ref{sec:interaction-foundations} \citep{Venkatesh2023Gaussian} tractable for continuous multimodal regression, yielding an explicit $R/U/S$ weighting interpretable against domain knowledge (redundancy dominates CT-slice regression; synergy dominates the vision--text pair in sentiment).

The same PID-typed disentanglement recurs across clinical deployments. \citet{Zhang2024Prototypical}'s PIBD is representative: for cancer survival prediction from gigapixel pathology images and genomic pathways under weak bag-level supervision, a Prototypical Information Bottleneck approximates the intractable posterior with per-risk-band prototypes, then splits the selected features into modality-common and modality-specific parts by minimizing their CLUB-estimated mutual information; across five TCGA datasets it achieves the best concordance index, and intervening on the prototypes collapses performance. DrFuse \citep{Yao2024DrFuse} applies the same factorization to EHR-plus-chest-X-ray fusion under missing modalities (Table~\ref{tab:pid-methods}); M-IDoL \citep{Liu2026IDoL} instead disperses modalities into separable mixture-of-experts subspaces via entropy objectives, for medical foundation-model pretraining across 21 tasks. Adjacent but not a formal PID method, SFL-Net \citep{Chopra2026Interpretable} learns a source-factorized (shared, modality-specific, complementary) quantized latent for MRI-to-PET synthesis; the factorization is PID-inspired, and the authors state explicitly that it does not estimate PID atoms.

Beyond the clinic, PID typing targets missing modalities, noisy fusion, label scarcity, and communication topology. \citet{Wang2025Multimodal}'s MRdIB first filters noise via a Multimodal Information Bottleneck, then layers one PID-derived objective per atom -- maximizing unique information and synergy while minimizing redundancy. \citet{Nguyen2025Robult} ground their design directly in the bivariate PID identity, targeting redundancy and per-modality unique information under missing modalities and label scarcity. \citet{Shi2026PIDb}'s PARSE carries the typing into decentralized federated learning via ``feature fission'': each agent factorizes its per-modality latent into redundant, unique, and synergistic slices, then exchanges only redundant slices over per-modality subgraphs and shares synergy-head parameters only among agents holding the same modality set, showing PID typing governs \emph{which parameters to communicate} as naturally as which features to keep. Finally, \citet{Yang2026Information}'s DMIL exploits sample-level variation in $R/U/S$ via a two-stage variational decomposition guided by an MI-to-performance lower bound that assumes deterministic encoding and a deterministic interaction composition. \citet{Dong2023SimMMDG}'s SimMMDG proves that, under exact feature equality across modalities (a condition stronger than CLIP's shared-space similarity alignment), the optimal cross-entropy risk is worse than the raw-input optimum by at least the gap $\Delta_p:=|I(X_1;Y)-I(X_2;Y)|$ between the modalities' individual predictive information -- a domain-generalization analogue, under different assumptions, of \citet{Jiang2023Understanding}'s modality-gap bound (\S\ref{sec:foundation-diagnosis}).

\subsection{Interaction-Aware Mixture-of-Experts Architectures}
\label{sec:interaction-moe}

The most direct architectural payoff is to stop measuring redundancy, uniqueness, and synergy post hoc and instead build one expert per interaction type, routing each sample by the interaction it exhibits -- a move that presupposes the typing \S\ref{sec:interaction-foundations} and \S\ref{sec:interaction-measurement} supply.

The cheapest instantiation, \citet{Yu2024MMOE}'s Mixtures of Multimodal Interaction Experts (MMOE), is motivated by a striking observation: a single fusion model reaches roughly 89\% F1 on sarcasm detection when modalities are redundant but collapses to roughly 24\% F1 when the sarcastic intent is synergistic. Rather than estimate PID atoms, MMOE types each training point via a cheap prediction-discrepancy heuristic and trains one expert per interaction type. \citet{Xin2025MoE}'s I2MoE makes this end-to-end and PID-grounded: four experts instantiated directly from the taxonomy, trained with a weakly supervised interaction loss that perturbs one modality at a time, plus a learned re-weighting model producing sample-level soft importance weights -- so expert specialization is jointly learned, not statically pre-partitioned. \citet{Yu2026PathMoE} show the architecture transfers to data-scarce pediatric brain tumor subtyping from whole-slide images, pathology-report text, and nuclei-level cell graphs: PathMoE extends I2MoE's expert structure with the cell-graph modality and a fifth expert, and its per-sample gating weights attribute diagnostic corrections to specific modality interactions in ways neuropathologists judged clinically plausible.

\emph{Additional surveyed methods.} The interaction-typing literature further includes pre-PID multi-view clustering and disentanglement \citep{Yan2024Cross,Yan2024Multi,Yan2024Differentiable,Zhang2025Disentanglement}, foundational estimator and normative studies \citep{Wollstadt2023Rigorous,Ehrlich2023Measure}, further quantification and attribution frameworks \citep{Singh2026SPICE,Halder2025Formalizing,Hu2022SHAPE,Dissanayake2025Quantifying,Chen2026PrismNet}.

\section{Interpretability, Efficiency, and Diagnosis in the Foundation-Model Era}
\label{sec:foundation}

Sections~\ref{sec:align} through~\ref{sec:interaction} traced how MI, IB, and PID shape multimodal representations from scratch. This section applies the same frameworks at the opposite end of the pipeline: rather than governing pretraining, MI, IB, and PID prune, tokenize, route, attribute, and diagnose pretrained models -- some as post-hoc probes on frozen vision--language models (VLMs), MLLMs, omni-modal LLMs, and diffusion models, others intervening during lightweight adaptation, routing, tokenization, or data selection. Here the frameworks converge on a single class of object (Table~\ref{tab:foundation-methods}), and the choice among them becomes a choice of question rather than one dictated by the challenge.

\begin{table*}[t]
\centering
\footnotesize
\caption{Representative foundation-model-era methods of Section~\ref{sec:foundation}. ``Operation'': the role the information-theoretic tool plays on a pretrained model, as a post-hoc probe or during lightweight adaptation.}
\label{tab:foundation-methods}
\begin{tabularx}{\textwidth}{@{}m{3.4cm}@{\hspace{5pt}}>{\centering\arraybackslash}m{0.9cm}@{\hspace{5pt}}X@{\hspace{5pt}}m{5.2cm}@{\hspace{5pt}}>{\centering\arraybackslash}m{1.2cm}@{}}
\toprule
Method & Frmwk. & Operation & Target model(s) & Subsec. \\
\midrule
MI-Pruner \citep{Li2026Pruner} & MI & Prune visual tokens (PMI selection) & LLaVA-1.5, Qwen2/2.5/3-VL, Video-LLaVA & \S\ref{sec:foundation-efficiency} \\
MERGE \citep{Han2026Massively} & PID & Route experts by temporal $R/U/S$ & Massively multimodal sensor models & \S\ref{sec:foundation-efficiency} \\
MINT \citep{Shan2025MINT} & PID & Group instruction-tuning data by interaction type & MLLMs on HEMM (18 datasets) & \S\ref{sec:foundation-efficiency} \\
\midrule
M2IB \citep{Wang2023Visual} & IB & Attribution / saliency & CLIP-style models & \S\ref{sec:foundation-interpretability} \\
\citet{Wu2026How} & PID & Layer-wise trajectory analysis & LLaVA-1.5/1.6 & \S\ref{sec:foundation-interpretability} \\
\midrule
\citet{Billa2026Modality} & MI & Diagnose modality collapse (GMI bound) & 5 speech / vision models & \S\ref{sec:foundation-diagnosis} \\
\citet{Jiang2023Understanding} & MI & Diagnose + regularize modality gap & CLIP / ALBEF pretraining & \S\ref{sec:foundation-diagnosis} \\
\bottomrule
\end{tabularx}
\end{table*}

\subsection{Efficiency: Pruning, Tokenization, and Routing at Scale}
\label{sec:foundation-efficiency}

A large multimodal foundation model spends compute uniformly across visual tokens, shared latent codes, and expert parameters, regardless of each one's task-relevant information. Four papers attack this budget problem -- what to keep, compress, or route -- at increasing scale, from individual tokens to whole datasets.

At the token level, \citet{Li2026Pruner}'s MI-Pruner reframes visual-token pruning as pointwise mutual information between projected visual and textual embeddings, casting selection as a monotone submodular maximization with a $(1-1/e)$ guarantee under the paper's conditional-independence assumption. This attention-free criterion retains only 5.57\% of Video-LLaVA tokens at 93.19\% relative accuracy, matching or beating attention- and diversity-based pruning.

One layer up, \citet{Tang2026InfoTok}'s InfoTok regularizes the shared visual tokenizer of a unified understanding-and-generation MLLM as an IB trade-off -- minimizing redundant input information $I(\tilde Z;X_{\mathrm{img}})$ (the raw image) while maximizing sufficiency $I(\tilde Z;Y^{\mathrm{GT}})$ and alignment $I(\tilde Z;T)$ -- improving understanding and generation benchmarks simultaneously with no extra data.

A third lens governs where to send a token rather than what to keep. \citet{Han2026Massively}'s MERGE extends PID to temporally lagged multi-sensor fusion, decomposing the interaction at each lag $\tau$ into redundancy $R(\tau)$, uniqueness $U_1/U_2(\tau)$, and synergy $S(\tau)$, then steering an RUS-aware Mixture-of-Experts router toward interaction-specialized experts with domain-interpretable routing.

The same PID vocabulary reappears at dataset granularity. \citet{Shan2025MINT}'s MINT computes a dataset-level redundant/unique/synergistic score from unimodal--multimodal prediction agreement, clusters instruction-tuning datasets into interaction-coherent groups, and fine-tunes each separately, beating unselective multi-task tuning by double-digit margins on HEMM.

\subsection{Interpretability and Attribution for Pretrained Multimodal Models}
\label{sec:foundation-interpretability}

A second cluster uses information theory to make a foundation model explicable, where the IB--PID contrast is sharpest: IB-based attribution asks \emph{how much} information flows through a bottleneck to an output, while PID-based attribution asks \emph{what kind} -- redundant, unique to one modality, or synergistic.

The IB line begins with M2IB \citep{Wang2023Visual}, which adapts supervised IB attribution to the label-free contrastive setting of CLIP-style models using the paired modality's embedding as a proxy target, outperforming GradCAM and attention-based attribution. \citet{Zhu2025Narrowing}'s NIB removes M2IB's stochastic noise and sensitive $\beta$ with a deterministic global-scalar bottleneck whose total feature importance provably equals the mutual information lost, improving text interpretability by nearly 59\% over state-of-the-art attribution baselines (including M2IB) while running about 64\% faster.

The PID line asks a categorically different question of the same black boxes. \citet{Zawar2024DiffusionPID}'s DiffusionPID extends a pixel-wise MI framework for diffusion models to a full PID over prompt phrases and generated pixels: on Stable Diffusion 2.1, redundancy maps expose gendered and ethnic occupational biases, synergy maps localize homonym disambiguation, and redundancy-guided interventions remove words where MI- and attention-based maps cannot. \citet{Amit2025Quantifying} bring the same decomposition to training-free VLM modality attribution via an Iterative Proportional Fitting / Sinkhorn estimator, finding that fusion direction systematically biases which modality appears dominant -- a confound outcome-driven metrics miss.

Two studies extend the lens from snapshot to trajectory. \citet{Wu2026How} track redundant, vision-unique, language-unique, and synergistic components layer-by-layer inside LLaVA-1.5/1.6, finding a three-stage ``modal transduction'' process that ends at 82\% language-unique versus 6\% vision-unique and under 2\% synergy, validated causally via attention knockout. \citet{Xiu2026Comprehensive} scale this across 26 open-source large vision--language models (LVLMs) and show synergy emerges during visual instruction tuning rather than earlier alignment pretraining.

\subsection{Diagnosing Failure Modes: Modality Collapse and the Modality Gap}
\label{sec:foundation-diagnosis}

A third cluster explains why pretrained foundation models fail -- collapsing onto one modality, misaligning when they should not -- and mostly fixes the failure with an IB-style regularizer.

\citet{Billa2026Modality} frames modality collapse -- an MLLM failing on tasks like emotion detection or counting though the needed information is present -- as an LLM decoder whose accessible information is bounded by Generalized Mutual Information (GMI). Under shared-marginal, local-Lipschitz, and bounded-region assumptions, the paper's GMI-Wasserstein bound caps the accessible-information drop by the decoder's Lipschitz sensitivity times the Wasserstein-1 distance between modal and text representations; the author notes the bound can be loose in the high-mismatch regime. Across five speech and vision models, linear probes confirm non-text information survives encoding while a targeted LoRA fine-tune resolves the failure, so an objective-side fix suffices. \citet{Jiang2023Understanding} prove via the MI chain rule and data-processing inequality that two perfectly aligned encoders incur downstream error exceeding the raw inputs by at least the gap between each modality's label-predictive information, making contrastive pretraining's implicit zero-modality-gap goal provably suboptimal; their three regularizers improve CLIP- and ALBEF-style pretraining.

Two IB-based papers address the erosion of pretrained visual competence under text-heavy instruction tuning. \citet{Wu2025Mitigating} frame ``visual forgetting'' as IB compression of the vision channel, using the effective rank of visual tokens as an information proxy and countering it with Modality-Decoupled Gradient Descent. \citet{Oh2025Visual}'s Vittle instead inserts a learnable Gaussian bottleneck with a variational lower bound of the IB objective, improving distribution-shift robustness across 45 datasets and 30 shift scenarios. \citet{Zhang2025Vision}'s VLV turns a bottleneck into a virtue, using a frozen text-to-image diffusion decoder as an implicit IB regularizer so that a lightweight captioner matches GPT-4o-competitive quality at roughly three orders of magnitude lower training cost.

This subsection's vocabulary is drawn almost entirely from MI and IB -- a conspicuous asymmetry relative to \S\ref{sec:foundation-interpretability}, where PID contributed roughly half the surveyed work. PID distinguishes redundant, unique, and synergistic contributions, so it would seem well suited to explaining \emph{why} a model collapses onto one modality; yet none of these papers apply PID to modality collapse or the modality gap.

Taken together, these methods diagnose and economize a model whose objective is already fixed; what remains absent is a typed PID reward embedded directly in foundation-model pretraining, the tension Section~\ref{sec:challenges} takes up.

\emph{Additional surveyed methods.} Further foundation-model studies include a controlled from-scratch scaling law \citep{Tong2026Beyond}, inference-time knowledge selection \citep{Hwang2025BOTTLEHUMOR}, behavioral cross-modal consistency testing \citep{Dagan2025CAST}, reward-model reliability diagnosis \citep{Huang2024Dark}, and the describe-then-generate information bottleneck \citep{Kodathala2025Describe}.

\section{Synthesis: A Unifying Coordinate System}
\label{sec:synthesis}

Sections~\ref{sec:align}--\ref{sec:foundation} have shown, challenge by challenge, how MI, IB, and PID relate in practice: alignment methods that bolt a compression term onto InfoNCE (\S\ref{sec:align-ib}), fusion architectures that call themselves MI-based yet reinvent IB's sufficiency-versus-compression trade-off, \citet{Federici2020Learning} most influentially (\S\ref{sec:robust-hybrid}), and disentanglement methods that push a binary shared/private split exactly to where PID's four-term decomposition becomes necessary (\S\ref{sec:interaction-preformal}--\S\ref{sec:interaction-foundations}). This section states those three relationships once, precisely, giving Sections~\ref{sec:challenges}--\ref{sec:conclusion} a single formal reference, and closes by mapping the survey's three motivating principles onto the conceptual progression the earlier sections demonstrated concretely.

\paragraph{A unified objective: the Generalized Multimodal Information Lagrangian} The three relationships stated below are, on their own, three separate prose claims about pairs of frameworks. This paragraph and the two that follow state a single object of which all three -- and the training objectives used throughout Sections~\ref{sec:align}--\ref{sec:interaction} -- are exact or approximate special cases. The two statements it rests on are elementary but load-bearing -- one restates the PID definition for learned encoders, the other the data-processing inequality in PID notation -- and it is their combination that fixes the coordinate system, not either result in isolation. Consider two modality-specific stochastic encoders $Z_1=f_1(X_1,\epsilon_1)$ and $Z_2=f_2(X_2,\epsilon_2)$ whose noise variables $\epsilon_1,\epsilon_2$ are exogenous -- independent of each other and of $(X_1,X_2,Y)$ -- the standard reparameterized setup of a per-modality (two-tower) stochastic encoder. Jointly fused IB architectures, which bottleneck a single $Z=g(X_1,X_2)$, satisfy this factorization only across their per-modality encoding stage. Exogeneity implies the familiar per-modality bottleneck Markov chains $Y - X_i - Z_i$ ($Z_i$ can carry information about $Y$ only through $X_i$), and also the joint channel factorization $p(z_1,z_2\,|\,x_1,x_2,y)=p(z_1\,|\,x_1)\,p(z_2\,|\,x_2)$ -- a factorization that guarantees the joint Markov chain the Aggregate Data-Processing Inequality below needs; the per-modality chains alone would not suffice there. Because the Williams--Beer decomposition \citep{Williams2010Nonnegative} (\S\ref{sec:interaction-foundations}) is defined for any pair of sources and one target, not only for the raw modalities $X_1,X_2$, it applies unchanged to the \emph{learned} pair $Z_1,Z_2$:

\textbf{Bottleneck PID Identity.} For any encoders $Z_1,Z_2$ and target $Y$,
\[
I(Z_1,Z_2;Y) \;=\; R_Z + U_{1,Z} + U_{2,Z} + S_Z,
\]
where $(R_Z,U_{1,Z},U_{2,Z},S_Z)$, all $\geq 0$, is the PID of $Y$ with respect to $(Z_1,Z_2)$ under a redundancy measure that induces a valid non-negative PID, constructed exactly as $R,U_1,U_2,S$ are constructed for $(X_1,X_2)$ in \S\ref{sec:interaction-foundations}. The identity is exact and requires no assumption beyond $R_Z,\dots,S_Z$ being a valid PID of $(Z_1,Z_2)\to Y$ -- it is not a property of any one estimator or architecture, but a consequence of applying the same decomposition one level downstream of where Section~\ref{sec:interaction} usually applies it. The decomposition is stated for the encoder \emph{pair} $(Z_1,Z_2)$ before fusion; a method that collapses them into a single fused vector $Z=g(Z_1,Z_2)$ inherits it only up to the information fusion discards, since $I(Z;Y)\leq I(Z_1,Z_2;Y)$ with equality when fusion is information-lossless. Read at the level of the pair, the identity says what an IB sufficiency term is made of: the joint task information a fusion objective in Section~\ref{sec:robust} seeks to retain is always a sum of four non-negative, qualitatively distinct quantities, and an untyped IB reward cannot see which one it is paying for. The identity separates interaction \emph{types}; it does not by itself distinguish a causal unique contribution from a dataset shortcut, which an observational PID may equally count in $U_1$ -- the shortcut failures of \S\ref{sec:robust-noise} therefore call for causal or interventional tools beyond PID.

\paragraph{The GMIL and its four corners} The Bottleneck PID Identity makes it possible to write down one Lagrangian that exactly nests the two-tower per-modality bottleneck objectives surveyed here (typed or untyped equal-reward alike) and approximately relates the rest, through the target substitution made precise below and the fused-bottleneck reduction described there. The surveyed objectives can be collected into one parameterized form, the \emph{Generalized Multimodal Information Lagrangian}
\[
\begin{aligned}
\mathcal{L}_{\mathrm{GMIL}} \;=\; &-\alpha_1 I(X_1;Z_1) - \alpha_2 I(X_2;Z_2)\\
&+ \beta_R R_Z + \beta_{U_1} U_{1,Z} + \beta_{U_2} U_{2,Z} + \beta_S S_Z,
\end{aligned}
\]
maximized over the encoder parameters, with compression costs $\alpha_1,\alpha_2\geq 0$ and typed rewards $\beta_R,\beta_{U_1},\beta_{U_2},\beta_S\in\mathbb{R}$. Four qualitatively different training objectives already surveyed here map onto four corners of this one objective: untyped MI maximization (Section~\ref{sec:align}), multimodal IB (Section~\ref{sec:robust}), synergy-maximizing contrastive learning, and uniqueness-favoring disentanglement (Section~\ref{sec:interaction}). The mapping is approximate: label-free alignment reaches its corner only through a target substitution made precise below, and multimodal IB typically compresses a fused rather than per-modality representation. The Lagrangian has six coefficients but only their ratios matter, so it has five effective degrees of freedom, and its atoms are defined relative to a fixed redundancy measure inducing a valid non-negative PID. The GMIL as written is also strictly bivariate: multi-modality methods surveyed here (e.g., $K$-modality contrastive tuples and tri-modal PID extensions) lie outside it, and extending it beyond two modalities faces exactly the scaling barrier of \S\ref{sec:challenges-scaling}.

Two things follow. First, within this family IB (Section~\ref{sec:robust}) is PID-typed learning (Section~\ref{sec:interaction}) with the four rewards forced equal, and switching the compression cost off leaves the untyped reward $I(Z_1,Z_2;Y)$. MI-maximizing alignment (Section~\ref{sec:align}) reaches that corner only as an idealization: substituting a co-occurring modality for $Y$ yields $I(Z_1,Z_2;X_2)$ rather than CLIP's $I(Z_1;Z_2)$. The three frameworks are thus ordered by how many of the GMIL's coefficients they leave free, and the ``measurable $\to$ allocated $\to$ not monolithic'' progression reads as nested parameter constraints up to that substitution. Second, the synergy-only and uniqueness-only corners are not hypothetical: \citet{Chen2026SynGR} instantiate $\beta_S \gg \beta_R,\beta_{U_1},\beta_{U_2}$, while \citet{Hazarika2020MISA} and \citet{Nguyen2025Robult} are the closest occupants of the uniqueness-favoring corner. (\citet{Proca2022Synergistic}, by contrast, measure rather than optimize how synergy emerges under task pressure.) \citet{Liang2023Factorized} give the formal account of when that target substitution is safe (multi-view redundancy) and when it fails (task-relevant unique information) -- that failure, not the corner itself, is what makes Section~\ref{sec:interaction}'s typed rewards necessary.

\paragraph{Relation to prior unified objectives} The GMIL synthesizes convergent practice rather than proposing a new algorithm. \citet{Abdelaleem2023Deep} already unify VAEs, deep variational IB, deep variational CCA, CLIP, and Barlow Twins as one $\mathcal{L}=I_{\text{encoder}}-\beta I_{\text{decoder}}$ family; the GMIL's only structural addition is to replace the scalar decoder reward with the \emph{typed} vector $(\beta_R,\beta_{U_1},\beta_{U_2},\beta_S)$ the Bottleneck PID Identity licenses. Modest as that change is, it is exactly what a scalar decoder reward cannot express: a synergy-favoring corner ($\beta_S$ large) and a redundancy-favoring corner ($\beta_R$ large) are indistinguishable to a family that sees only the total $I(Z_1,Z_2;Y)$, yet demand opposite designs -- which is why alignment, robust fusion, and interaction typing occupy distinct regions here rather than coinciding. The typed objective is not hypothetical: \citet{Wang2025Multimodal}'s MRdIB (\S\ref{sec:interaction-representation}) is an (at least approximate) GMIL instance predating our statement of it. As an account of \emph{existing} objectives the GMIL is thus expository -- but as a coordinate system over the family it defines, it also localizes the regions no surveyed method occupies, as we show next.

\paragraph{The unoccupied corners: method families the coordinate system points to} These four occupied corners share a revealing feature: all either leave the four typed rewards \emph{equal} (untyped MI, IB) or push a \emph{single} atom's reward far above the rest (synergy, uniqueness). Large tracts of the space -- corners rewarding redundancy on its own, suppressing synergy with a \emph{negative} weight, tying per-modality compression $\alpha_i$ to unique-information content, or demanding a typed reward \emph{and} heavy compression at once -- are occupied by no method in our screened corpus -- a census claim under the stated search protocol (Section~\ref{sec:intro}), not a proof of non-existence. Table~\ref{tab:gmil-predicted} reads each as a concrete method family the coordinate system points to: a parameter regime, the capability it would confer, and the closest surveyed work. Supplementary Table~S2 places representative surveyed methods in this coordinate system, projected onto compression strength and typed-reward composition. The placement makes the tier structure concrete: the exact tier is populated only at the untyped equal-reward point -- exactly in form by the two-tower multi-view IB family \citep{Federici2020Learning,Abdelaleem2023Deep,Wen2024MVEB}, whose label-free targets still require the substitution above, and exactly outright by supervised distributed IB \citep{Zhou2025Distributed}; \emph{no surveyed method occupies the typed exact tier}, which the typed corners reach only through surrogate atoms or fused bottlenecks. The last row is not a corner of the GMIL as written, whose atoms are population-level scalars, but a proposed extension requiring pointwise or conditional atoms $r(x_1,x_2,y)$, $u_1(\cdot)$, $u_2(\cdot)$, $s(\cdot)$ with realization-dependent rewards $\mathbb{E}[\beta_R r+\beta_{U_1}u_1+\beta_{U_2}u_2+\beta_S s]$; like pointwise MI, local atoms can be negative, which any such loss must handle. Each defines a formally specified candidate objective family -- realizable in practice only insofar as the rewarded atoms admit a tractable, stable estimator or surrogate, the very gap \S\ref{sec:challenges-estimation} documents -- and at least one is the formal statement of a gap Section~\ref{sec:challenges} arrives at from the literature side -- the two formulations pick out the same absence.

\begin{table*}[t]
\centering
\footnotesize
\caption{Unoccupied corners of the GMIL, plus one proposed extension (last row): for each, a parameter regime no surveyed method targets as its primary objective, the capability it would confer, and the closest existing work.}
\label{tab:gmil-predicted}
\renewcommand{\arraystretch}{1.25}
\begin{tabularx}{\textwidth}{@{}m{2.9cm}@{\hspace{5pt}}m{3.4cm}@{\hspace{5pt}}X@{\hspace{5pt}}m{4.0cm}@{}}
\toprule
Unoccupied regime & Predicted method family & Why it should exist & Closest surveyed work \\
\midrule
$\beta_R \gg \beta_{U_1},\beta_{U_2},\beta_S$ (all $>0$) & Redundancy-\emph{maximizing} pretraining & Rewards information each modality carries on its own, so a dropped modality costs less wherever such information exists (it cannot help where $I(X_i;Y)=0$, as in XOR): \emph{proactive} missing-modality robustness & \S\ref{sec:robust-missing} \emph{reconstructs} lost information reactively instead \\
$\alpha_1\neq\alpha_2$, tied to $U_1$ vs.\ $U_2$ & Interaction-typed asymmetric compression & Compress the mostly redundant modality, preserve the one carrying unique information & OMIB \citep{Wu2025Learning} weights modalities by task information, not by the unique atom \\
$\beta_S<0$ (synergy \emph{suppressed}) & Synergy-averse fail-safe fusion & Discourage reliance on fragile joint-only information where it is unacceptable (safety-critical, intermittent sensors); suppression alone does not convert synergy into redundancy & None in the corpus; synergy is only ever maximized, never avoided \\
$\beta_S \gg$ others \emph{and} $\alpha$ large & Compression-robust synergy learning & Retain synergy under a hard bandwidth or token budget & SynIB \citep{Kontras2026SynIB} targets synergy under an IB framing but adds no explicit compression term or bandwidth budget \\
Pointwise GMIL with per-sample typed rewards (extension) & Interaction-typed foundation-model pretraining & Put the typed reward \emph{inside} the pretraining loss, not a post-hoc probe or bolt-on router & MoE routing (\S\ref{sec:interaction-moe}) types per sample only in small models; MINT \citep{Shan2025MINT} routes data, not the objective \\
\bottomrule
\end{tabularx}
\end{table*}

\paragraph{From coordinates to a heuristic design guide} The same coordinate system can also guide a practitioner toward which \emph{existing} method to try first. Because a method's corner is fixed by the interaction profile it targets, choosing a method is choosing where in parameter space a task lives -- an empirical question \S\ref{sec:interaction-measurement}'s quantification tools answer before training. Table~\ref{tab:decision} turns this into a heuristic design guide: the task's measured dominant atom and deployment constraint suggest candidate regimes to test empirically -- hypotheses for model selection, not optimality guarantees.

\begin{table}[t]
\centering
\footnotesize
\caption{A heuristic design guide: a task's measured dominant atom (\S\ref{sec:interaction-measurement}) suggests a parameter regime and candidate method families; the mappings are hypotheses, not optimality guarantees.}
\label{tab:decision}
\setlength{\tabcolsep}{4pt}
\renewcommand{\arraystretch}{1.25}
\begin{tabularx}{\columnwidth}{@{}m{1.8cm}@{\hspace{4pt}}m{1.8cm}@{\hspace{4pt}}X@{}}
\toprule
Dominant atom & Constraint & GMIL regime $\Rightarrow$ method family \\
\midrule
Redundancy & missing-modality risk & high $\alpha$, uniform $\beta$ $\Rightarrow$ IB / redundancy-robust fusion (Section~\ref{sec:robust}) \\
Uniqueness & --- & $\beta_{U_i}>0$, $\beta_R$ small $\Rightarrow$ PID disentanglement (\S\ref{sec:align-pid}, \S\ref{sec:interaction-representation}) \\
Synergy & clean modalities & $\beta_S \gg$ others $\Rightarrow$ synergy-max contrastive or interaction-typed MoE (\S\ref{sec:align-pid}, \S\ref{sec:interaction-moe}) \\
Unknown & decide pre-training & measure first $\Rightarrow$ PID-based model selection (\S\ref{sec:interaction-measurement}) \\
Any & frozen foundation model & no training $\Rightarrow$ post-hoc PID attribution (\S\ref{sec:foundation-interpretability}) \\
\bottomrule
\end{tabularx}
\end{table}

\paragraph{Compression can only shrink typed information in aggregate; only the term-by-term refinement is measure-dependent} The GMIL's compression terms and typed rewards are not independent. Because the encoder noises are exogenous, the channel factorization above makes $(Z_1,Z_2)$ a stochastic function of $(X_1,X_2)$ alone, so $Y-(X_1,X_2)-(Z_1,Z_2)$ is a Markov chain and the data-processing inequality gives $I(Z_1,Z_2;Y)\leq I(X_1,X_2;Y)$. The inequality below is, in this sense, the data-processing inequality written in PID notation: because both sides are joint mutual informations, the aggregate bound holds for \emph{any} choice of redundancy measure, and its content is precisely that the total typed information cannot grow under compression. The exogeneity assumption is doing real work here: the per-modality chains $Y-X_i-Z_i$ by themselves do not imply the joint chain, since encoder noises correlated through $Y$ can make the \emph{pair} $(Z_1,Z_2)$ strictly more informative about $Y$ than $(X_1,X_2)$ is -- an XOR-style construction that the factorization rules out. Combining this with the Bottleneck PID Identity and the original Williams--Beer identity for $(X_1,X_2)$ yields:

\textbf{Aggregate Data-Processing Inequality.}
\[
R_Z + U_{1,Z} + U_{2,Z} + S_Z \;\leq\; R + U_1 + U_2 + S.
\]

Compression can only ever destroy typed information in aggregate, never manufacture it: an encoder cannot increase the aggregate sum of the four PID atoms beyond the raw joint task information. What it does \emph{not} say is which atom compression destroys first. Whether an aggressively compressed $Z$ preferentially loses synergy, redundancy, or one modality's unique contribution is a term-by-term claim that requires the chosen PID measure to satisfy an additional monotonicity property under degraded sources, and \S\ref{sec:interaction-foundations}'s survey of mutually incompatible estimators -- discrete Williams--Beer \citep{Williams2010Nonnegative}, copula-based continuous BROJA \citep{Pakman2021Estimating}, closed-form Gaussian \citep{Venkatesh2023Gaussian}, and flow-based Thin-/Flow-PID \citep{Zhao2025Partial} -- shows this is not guaranteed uniformly across the estimators this paper surveys. The gap between an aggregate bound provable from first principles and a term-by-term bound that depends on estimator choice is itself a precise, rather than rhetorical, statement of the open estimator-comparison problem taken up in Section~\ref{sec:challenges}.

\paragraph{From principles to a demonstrated chain} The GMIL encodes the survey's three principles as one relationship each. PID is a strict refinement of MI: the Williams--Beer identity \citep{Williams2010Nonnegative} is an exact partition, so a PID method optimizes a finer-grained version of the scalar an MI method reports whole. IB is built \emph{on} MI: its compression and prediction terms are the same estimators of \S\ref{sec:align-toolbox} \citep{Belghazi2018Mutual,Oord2018Representation,Cheng2020CLUB} asked to do two jobs at once \citep{Abdelaleem2023Deep}. And PID decomposes precisely IB's sufficiency term (the Bottleneck PID Identity); complementarily, \citet{Liang2023Factorized} show that standard contrastive learning, which maximizes cross-modal shared information, retains less task information as task-relevant unique information grows -- the gap the GMIL's typed rewards close.

\section{Open Challenges and Future Directions}
\label{sec:challenges}

Placing MI-, IB-, and PID-based treatments of the same problem side by side makes visible where a framework has simply not yet been tried. This section collects three such gaps, the first spanning both estimation and cross-framework evaluation.

\subsection{Estimator Intractability at Scale, Especially for PID}
\label{sec:challenges-estimation}

\looseness=-1 Every framework surveyed here depends on estimating mutual-information-type quantities from finite, high-dimensional samples; PID is the sharpest instance. \S\ref{sec:interaction-foundations} traces a decade of work devoted to making the Bertschinger et al.\ optimization tractable, from \citet{Pakman2021Estimating}'s copula-based variational bound for continuous data, to \citet{Venkatesh2023Gaussian}'s closed-form restriction to the jointly Gaussian case, to \citet{Zhao2025Partial}'s Thin-PID and Flow-PID solvers that extend an exact Gaussian optimum to arbitrary continuous modalities via normalizing flows. This sequence of increasingly specialized fixes shows that no general-purpose, sample-efficient PID estimator yet exists for the unrestricted continuous, high-dimensional case. Part of this difficulty is fundamental: \citet{McAllester2020Formal} prove that any distribution-free high-confidence lower bound on mutual information from $N$ samples is capped at $O(\ln N)$, a cap no amount of estimator engineering escapes. The same pressure shows up in mainstream multimodal ML: \citet{Liang2023Quantifying}'s CVX and BATCH estimators, the field's central hub, scale poorly with category count -- a gap \citet{Yang2025Efficient}'s pointwise LSMI estimator closes -- and \citet{Choi2026ICYM} show a correctly estimated PID can still be systematically biased under realistic missing-at-random shifts between training and deployment. MI and IB estimation face a milder version of the problem -- \citet{Tian2023Variational}, for instance, eliminate MI estimation from a multi-view IB objective via the chain rule -- but PID's four simultaneously estimated, non-negative terms compound the difficulty. Compounding this, MI-, IB-, and PID-based methods targeting the identical problem -- for instance the MI-diagnostic \citep{Gao2025Asymmetric} and analytically derived IB \citep{Wu2025Learning} treatments of modality imbalance in \S\ref{sec:robust-imbalance} -- are compared only narratively here, never on a common benchmark, because the three frameworks originate in sub-communities with different evaluation conventions. A standardized suite holding task, data, and architecture fixed while varying only the information-theoretic objective would separate genuine framework advantages from confounded tuning.

\subsection{Scaling PID Beyond Two or Three Modalities}
\label{sec:challenges-scaling}

\looseness=-1 PID's redundancy lattice is defined for an arbitrary number of sources in principle, but nearly every tractable estimator in \S\ref{sec:interaction-foundations} and \S\ref{sec:interaction-measurement} is derived and benchmarked for exactly two sources and one target. \citet{Wenderoth2025Measuring} build InterSHAP, a Shapley Interaction Index, as a substitute because it generalizes to any number of modality coalitions, whereas the scalable dataset-level PID estimators compared there are restricted to two modalities. The exceptions are recent and narrow: \citet{Han2026Massively}'s MERGE (\S\ref{sec:foundation-efficiency}) extends PID to a temporal, multi-source setting by decomposing interaction at each time lag rather than jointly across all sources; \citet{Fang2026Understanding}'s conditional ``Sensory PID'' extends the decomposition to tri-modal omni-models; and \citet{Liang2024Multimodal} derive bounds rather than exact estimates on synergy precisely because exact PID becomes unrecoverable once only unimodal labels plus unlabeled multimodal co-occurrence data are available -- a partial-observability problem that worsens with modality count. General, scalable PID estimation for four or more genuinely distinct modalities remains largely unsolved.

\subsection{From Post-Hoc Diagnosis to Training-Time Integration in Foundation Models}
\label{sec:challenges-foundation-gap}

\looseness=-1 Section~\ref{sec:foundation} surfaces a gap easy to state precisely. The MI- and IB-based diagnoses of foundation-model failure surveyed in \S\ref{sec:foundation-diagnosis} -- \citet{Billa2026Modality}'s Generalized-Mutual-Information account of modality collapse and \citet{Jiang2023Understanding}'s chain-rule proof that a zero modality gap is provably suboptimal -- are each followed by an IB-style fix. The surveyed corpus contains no comparable PID-based diagnosis or fix for these failure modes, despite PID being the framework built to distinguish whether a collapse reflects genuinely lost unique information versus a decoder treating unique visual and textual information as interchangeable. The PID-based work in Section~\ref{sec:foundation} -- \citet{Wu2026How}'s layer-wise trajectory analysis and \citet{Xiu2026Comprehensive}'s 26-model study, both in \S\ref{sec:foundation-interpretability} -- characterizes interaction type in models behaving normally, not in models actively failing, and none of it acts as a training-time signal. The closest steps toward closing this loop, \citet{Shan2025MINT}'s PID-guided grouping of instruction-tuning datasets and \citet{Han2026Massively}'s PID-guided expert routing (both \S\ref{sec:foundation-efficiency}), still operate at the level of data selection or routing, not inside the pretraining objective. Turning PID's redundancy/uniqueness/synergy vocabulary into a training-time regularizer for foundation-model pretraining, rather than a post-hoc probe, remains an open problem unaddressed anywhere in the corpus. This gap coincides with the proposed extension in the last row of Table~\ref{tab:gmil-predicted}, which the coordinate-system argument of Section~\ref{sec:synthesis} states formally.

Beyond these three gaps, two cross-cutting questions remain open. First, identifiability: for continuous modalities the estimated atoms depend on the chosen redundancy measure and estimator family (\S\ref{sec:interaction-foundations}), and the stability of a PID-guided training signal across those choices is essentially uncharacterized. Second, causality: an observational PID cannot distinguish a causal unique contribution from a dataset shortcut (Section~\ref{sec:synthesis}), so PID-guided training can reward exactly the artifact robust fusion tries to remove; combining typed-information objectives with interventional or counterfactual data is unexplored, as are the privacy and fairness implications of a unique atom that isolates a protected attribute.

\section{Conclusion}
\label{sec:conclusion}

This survey has organized 170 recent works (2018--2026), alongside 12 foundational references, within a unified information-theoretic framework rather than three separate literatures: MI, IB, and PID are successive refinements of one inquiry into inter-modal dependency -- how much two modalities depend on each other, how much of that dependency a representation should keep under compression, and what kind of dependency is preserved. Section~\ref{sec:align} showed MI maximization answering the first through cross-modal alignment; Section~\ref{sec:robust}, the information bottleneck adding an explicit compression term on top; Section~\ref{sec:interaction}, partial information decomposition refining IB's scalar prediction term into redundant, unique, and synergistic atoms. Scaling to multimodal foundation models (MM-FMs) is not a fourth type of dependency but the capstone at which measurability, allocation, and typing must be solved at once on already-pretrained models, and Section~\ref{sec:foundation} showed all three converging there, mostly as post-hoc or adaptation-time instruments rather than pretraining objectives. Section~\ref{sec:synthesis} anchored these relationships in a shared coordinate system -- the Bottleneck PID Identity, the Aggregate Data-Processing Inequality, and the Generalized Multimodal Information Lagrangian -- whose unoccupied parameter regions point toward model families not yet built in the surveyed literature, and Section~\ref{sec:challenges} marked the frontiers that remain: high-dimensional estimation tractability, the lack of cross-framework benchmarks, combinatorial explosion in multimodal PID, and the gap between post-hoc diagnosis and end-to-end training integration.

The field's trajectory across Section~\ref{sec:foundation} points toward where this inquiry is heading next: MI, IB, and PID are increasingly deployed not to design a multimodal model from scratch but to prune, attribute, and diagnose foundation models whose training objectives were fixed long before any of these tools were applied to them. Closing that gap -- building redundancy, uniqueness, and synergy directly into how the next generation of MM-FMs is pretrained, rather than only into how the current generation is explained after the fact -- marks the transition from post-hoc information analysis toward information-aware multimodal learning, and is the clearest direction in which dependency, compression, and synergy remain one unfinished question rather than three separate ones.

\bibliographystyle{IEEEtranN}
\bibliography{references}

\raggedbottom
\begin{IEEEbiography}[{\includegraphics[width=1in,height=1.25in,clip,keepaspectratio]{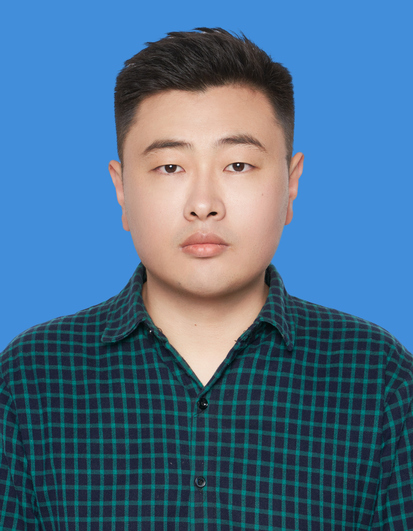}}]{Liangjian Wen} is currently an associate professor with the School of Computing and Artificial Intelligence, Southwestern University of Finance and Economics, Chengdu, China. He received the Ph.D. degree from the University of Electronic Science and Technology of China in July 2021. His research interests include machine learning, deep learning, representation learning, and self-supervised learning.
\end{IEEEbiography}

\begin{IEEEbiography}[{\includegraphics[width=1in,height=1.25in,clip,keepaspectratio]{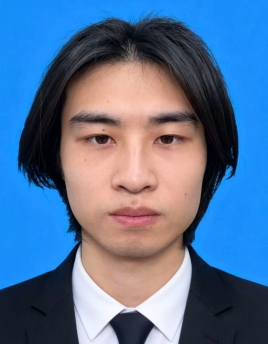}}]{Linjie Li} is currently pursuing the B.E. degree in Artificial Intelligence with the School of Computing and Artificial Intelligence, Southwestern University of Finance and Economics, Chengdu, China. His research interests include machine learning, deep learning, multimodal learning, and representation learning.
\end{IEEEbiography}

\begin{IEEEbiography}[{\includegraphics[width=1in,height=1.25in,clip,keepaspectratio]{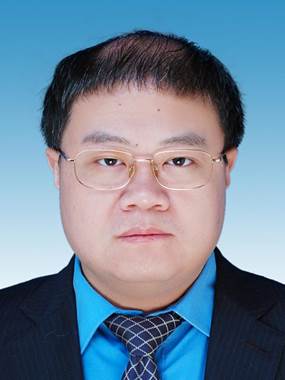}}]{Jiang Duan} received the B.Sc. degree in mechanical engineering from Southwest Jiaotong University, Chengdu, China, in 2001, the M.Sc. degree from the University of Derby, Derby, U.K., in 2002, and the Ph.D. degree from the University of Nottingham, U.K., in 2006. He is a professor at Southwestern University of Finance and Economics, China. He is a recognized national expert, the winner of the first Sichuan Outstanding Talent Award (The Highest Talent Award of Sichuan Province), the winner of the Sichuan Youth Science and Technology Award, and a standing committee member of the Sichuan Association for Science and Technology.\end{IEEEbiography}

\begin{IEEEbiography}[{\includegraphics[width=1in,height=1.25in,clip,keepaspectratio]{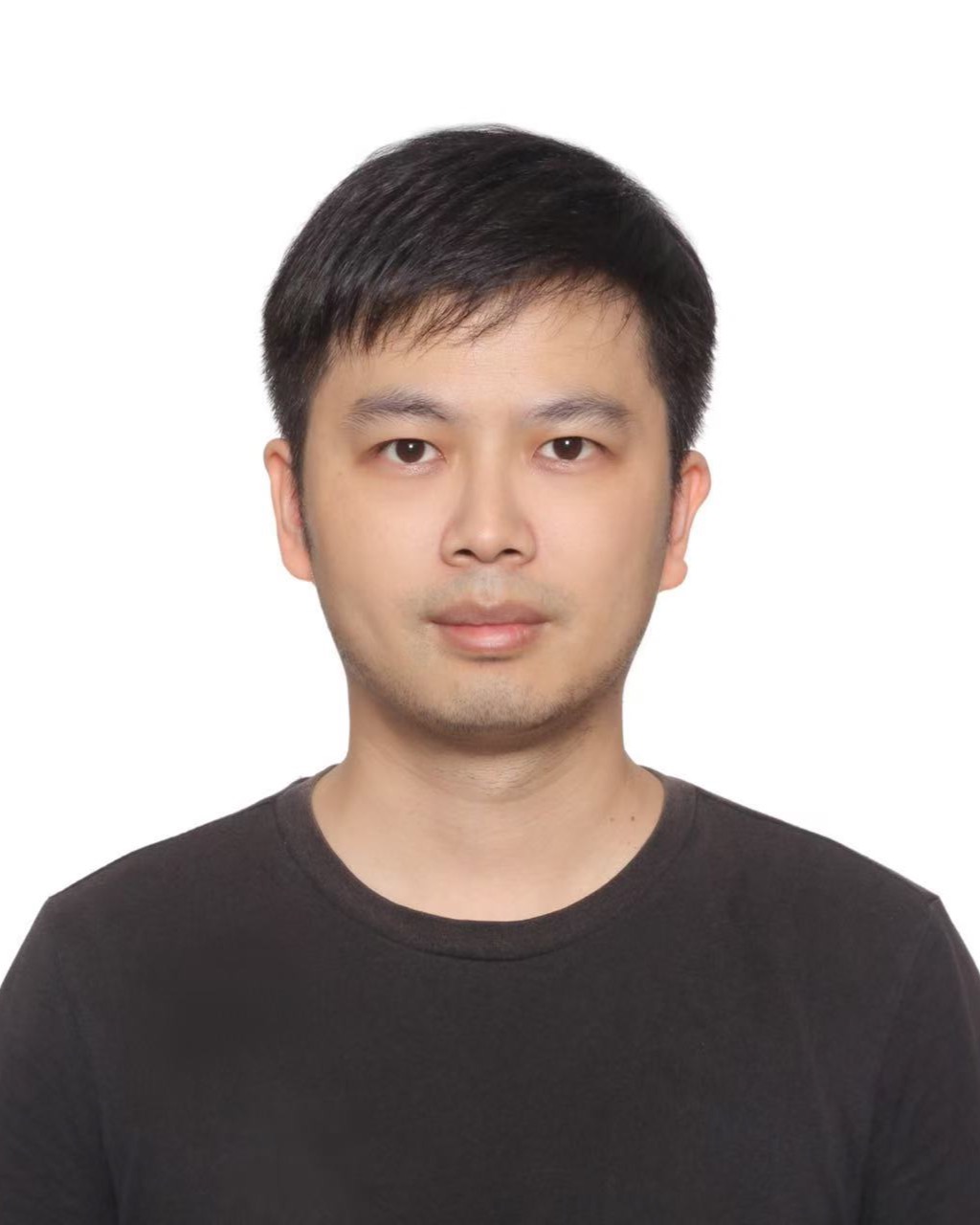}}]{Yong Dai} received the Ph.D. degree from the University of Electronic Science and Technology of China in 2021. He previously conducted research on large language models at Microsoft Research Asia and at Tencent AI Lab. Since May 2025, he has served as Head of Multimodal Large Models at the Beijing Humanoid Robot Innovation Center, leading research on embodied brain-cerebellum systems and world models.
\end{IEEEbiography}

\begin{IEEEbiography}[{\includegraphics[width=1in,height=1.25in,clip,keepaspectratio]{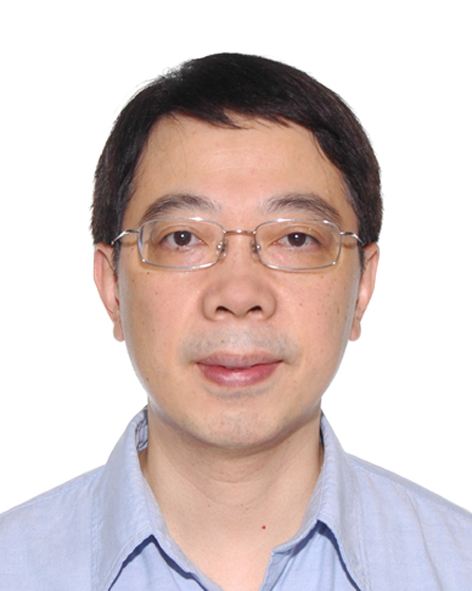}}]{Jianzhuang Liu} (Senior Member, IEEE) received the Ph.D. degree in computer vision from The Chinese University of Hong Kong in 1997. He is a professor at the Shenzhen Institutes of Advanced Technology, Chinese Academy of Sciences, and was a principal researcher at Huawei from 2012 to 2023. He has authored more than 200 papers in computer vision, image processing, machine learning, and multimodal AI.
\end{IEEEbiography}

\begin{IEEEbiography}[{\includegraphics[width=1in,height=1.25in,clip,keepaspectratio]{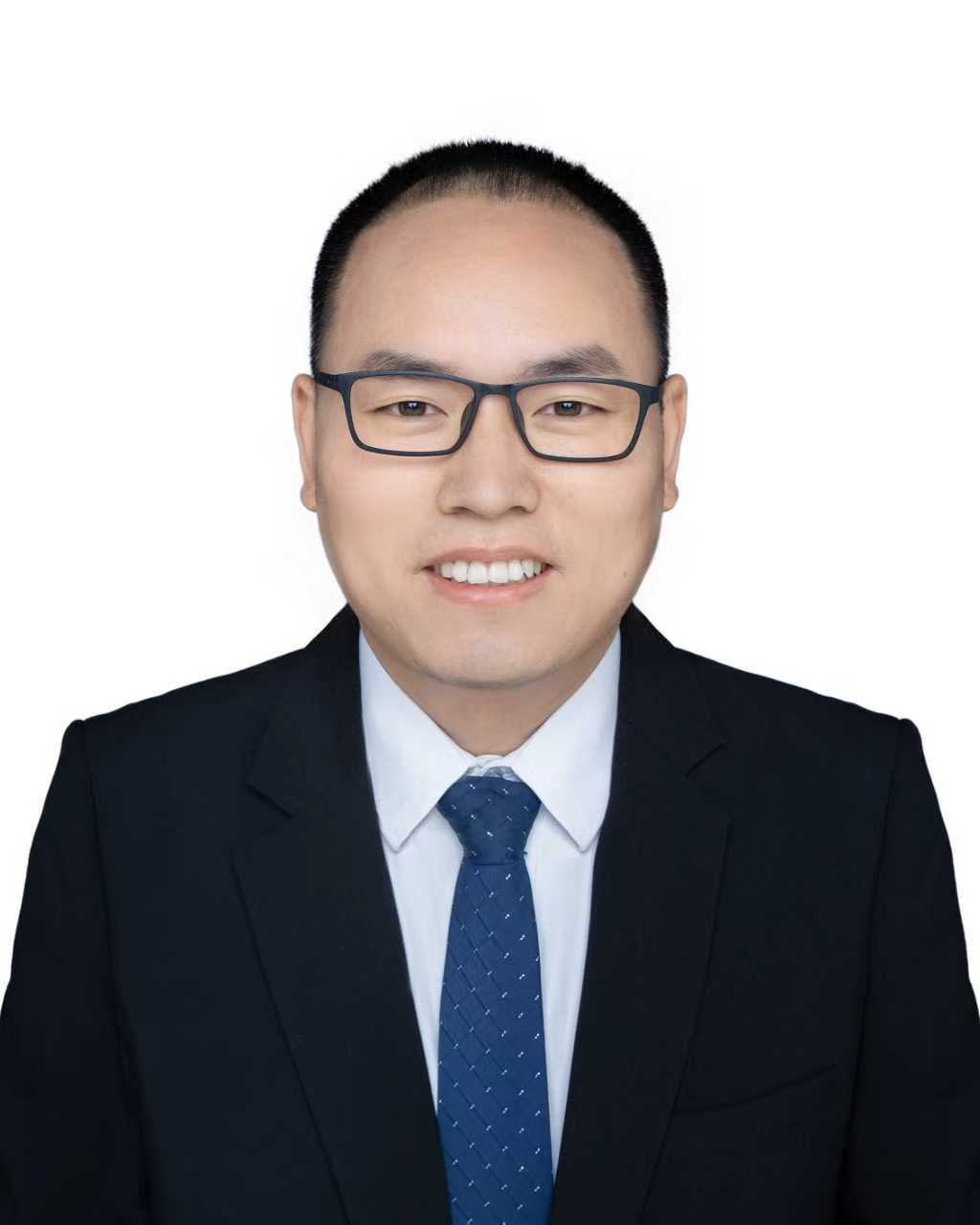}}]{Zhao Kang} received the Ph.D. degree in computer science from Southern Illinois University Carbondale, Carbondale, IL, USA, in 2017. He is currently a professor in the School of Computer Science and Engineering, University of Electronic Science and Technology of China, Chengdu, China. His research interests include graph machine learning and large language models. He serves as an Action Editor for Neural Networks.
\end{IEEEbiography}

\end{document}